\documentclass[a4paper,american,aps,citeautoscript,floatfix,longbibliography,pdftex,pra,
superscriptaddress,showpacs,preprint]{revtex4-1}

\usepackage{adjustbox}
\usepackage{enumerate,tensor}
\usepackage[normalem]{ulem}
\usepackage{natbib} 
\usepackage{hyperref}
\usepackage{amsmath,amssymb,amsmath}
\usepackage{graphicx}
\usepackage{dcolumn}
\usepackage{bm}
\usepackage{color}

\makeatletter
\def\paragraph{\@startsection{paragraph}{4}{10pt}{-1.25ex plus -1ex minus -.1ex}{0ex plus 0ex}{\normalsize\textit}}
\renewcommand\@biblabel[1]{#1}
\renewcommand\@makefntext[1]%
{\noindent\makebox[0pt][r]{\@thefnmark\,}#1}
\DeclareRobustCommand\onlinecite{\@onlinecite}
\def\@onlinecite#1{\begingroup\let\@cite\NAT@citenum\citealp{#1}\endgroup}
\def\tagform@#1{\maketag@@@{\ignorespaces#1\unskip\@@italiccorr}}
\let\orgtheequation\theequation
\def\theequation{(\orgtheequation)}
\makeatother

\begin{document}

\author{C. Chandre}
\affiliation{Aix Marseille Univ, CNRS, Centrale Marseille, I2M, Marseille, France}

\author{J. Pablo Salas}
\affiliation{\'Area de F\'{\i}sica, Universidad de la Rioja, 26006 Logro\~no, La Rioja, Spain}

\title{Formation of RbCs dimers using an elliptically polarized laser pulse} 

\date{\today}

\begin{abstract}
We consider the formation of RbCs by an elliptically polarized laser pulse. By varying the ellipticity of the laser for sufficiently large laser intensity, we see that the formation probability presents a strong dependence, especially around ellipticity $1/\sqrt{2}$. We show that the analysis can be reduced to the investigation of the long-range interaction between the two atoms.  The formation is mainly due to a small momentum shifts induced by the laser pulse. We analyze these results using the Silberstein's expressions of the polarizabilities, and show that the ellipticity of the field acts as a control knob for the formation probability, allowing significant variations of the dimer formation probability at a fixed laser intensity, especially in the region around an ellipticity of $1/\sqrt{2}$. 
\end{abstract}

\maketitle 

\section{Introduction}
The interaction of matter with the laser light is of fundamental importance in atomic and molecular physics, 
and in addition in current modern technology.
Indeed, an atom subjected
to intense laser fields is a good example of how
the complex interplay between the electrons-core Coulomb force and the
force exerted by the electric field leads to single, double and, in general,
multiple ionization processes (see, e.g., Refs.~\cite{A947,A652} and references therein).
On the other side, by means of strong laser fields, a feasible control of the alignment and orientation of molecules
is possible~\cite{A513,A477,A481,A484,A759,A747,A213,A949}. This issue is of great relevance in chemical
reactions because, in many situations, the
reaction rate is very sensitive to the relative orientation between the reactants~\cite{A939}.
In this sense, sophisticated experimental control schemes using
ultrashort laser pulses have made possible: i)
to select and manipulate particular reaction channels~\cite{A942}; ii) to
design specific femtosecond pulses to
maximize the yield
of the single ionization channel of a organometallic molecule, while the competing fragmentation 
channel was hindered~\cite{A943}; iii) to used few-cycle subfemtosecond pulses to break
an specific hydrogen bond in the deprotonation of a symmetric hydrocarbon~\cite{A937}; iv) to control
the fragmentation angular distribution of
photodissociation processes 
by means of an intense near-infrared laser light~\cite{A946}.

In all cases, the intense laser pulses induce
drastic changes in the configuration of the targets by allowing chemical reactions to occur
in a certain way. The energy brought in by the laser field is channeled along the various degrees of freedom of the target to trigger these changes in a very complex way.  Thence, the precise understanding
of the laser-driven processes is a prerequisite to the control of the outcomes of the reactions.
By changing the parameters of the laser field, the products of the laser-matter interactions are changing since the energy brought in by the laser flows differently along the different degrees of freedom of the target. Besides the intensity and the frequency of the laser, when using elliptically polarized laser
light, the ellipticity of the laser appears as a convenient additional parameter
because it can be changed continuously without an increase
of the energy brought in to the system. Examples where elliptically polarized laser fields have been used
can be found in molecular
alignment~\cite{A513}. More recently, intense few cycle elliptically polarized laser pulses are
playing an important role in high harmonic generation experiments~\cite{A944,A945}.

From the point of view of the classical dynamics, the nonlinear
nature of the matter-light interaction makes these systems
very interesting for classical studies. The use of nonlinear dynamics to study the
quantal world  of atoms and molecules has a long history.
For example, the response of atomic and molecular systems
to diverse external field configurations has been widely studied by
using classical dynamics
(see, e.g., Refs.~\cite{A19,A33,A50,A205,PRA2007,EJPD2007,A286,PLA2010,A635,A741,A652}). 
In many cases, those classical approaches were unrivaled to provide an intuitive
explanation of the quantum mechanical results (see, e.g., Refs.~\cite{A19,gutzwiller,A311,A585,A652,Blumel}).

Following a similar scheme to the one we used recently in Ref.~\cite{PRA2017}, in this paper
we use classical dynamics to study the formation of cold RbCs dimers driven by a strong
elliptically polarized laser pulse.
In this way, the creation yield of RbCs molecules is explored as a function of the laser parameters, namely  the ellipticity and the strength electric field.
Besides the kinetic terms and the potential energy between
the Rb and Cs atoms, the
rovibrational Hamiltonian of the system includes 
the interaction between the molecular polarizability and the laser pulse.
Furthermore, the Hamiltonian depends explicitly on time because the
laser pulse envelope is made of a ramp-up, a plateau and a ramp-down. Hence, the system depends
explicitly on time and the corresponding Hamiltonian has 
3+1$/$2 degrees of freedom. For an ensemble of initial conditions, the yield of the driven reaction 
is explored by computing the
formation probability as a function of the strength and the ellipticity of the laser field.
From these numerical calculations, we find that there is a complex and strong dependence of
the formation probability with respect to the ellipticity and the electric field strength.
Indeed, for increasing ellipticity and for low and intermediate laser field strengths, there is a smooth variation in
the formation probability, such that at at around an ellipticity value of $1/\sqrt{2}$, it abruptly increases.
For high laser field amplitudes, the formation probability is very small for laser ellipticity below $1/\sqrt{2}$, presenting
a peak at that value, such that, for larger values of $\epsilon$, the
formation probability saturates.
We notice that the duration of the pulse plays only a minor role since it can be absorbed in the field strength (as a renormalized field strength).
By assuming that the very small changes in the radial and the angular momenta of the dimer
induced by the laser pulse are the main responsible for the formation,  we use the 
long range terms of the potential energy curve and the molecular
polarizabilities to build a simplified two dimensional
Hamiltonian. This reduced Hamiltonian allows us to obtain
an analytic approximate expression for the final energy of the dimer after the laser pulse.
We use this approximate expression to explain the observed complex behavior of the formation probability.

The paper is organized as follows: In Sec.~\ref{sec:ham} we establish
the classical rovibrational Hamiltonian governing the dynamics of the RbCs dimer in the presence of an
elliptically polarized laser field. A thorough study of the critical points of the potential energy surface
of the system is also presented in that section. Section~\ref{Phase} is devoted to the analysis of
the phase space structures in the neighborhood of the dissociation threshold.
In Sec.~\ref{driving}, we compute numerically the formation probabilities as functions of the laser parameters, and
the results are analyzed by using a static approximation. Finally, the conclusions
are provided in Sec.~\ref{Conclusions}.

\section{Hamiltonian models} 
\label{sec:ham}
We use the Born-Oppenheimer approximation to study the dynamics of the RbCs
molecule in its $^1 \Sigma^+$ electronic ground state
subjected to a strong elliptically polarized laser pulse.
For the description of the problem we use an inertial reference frame
$\hat{\bf r}=(\hat{\bf x}, \hat{\bf y}, \hat{\bf z})$ with
the origin at the center of mass of the molecule. 
In the absence of the laser pulse, the two atoms of the molecule interact through
the potential ${\cal E}(R)$ and its Hamiltonian reads
\begin{equation}
\label{eqn:H0}
H_0(R,P_R,\theta,P_\theta,\phi,P_\phi)=\frac{P_R^2}{2\mu}+\frac{P_\theta^2}{2\mu R^2}+\frac{P_\phi^2}{2\mu R^2\sin^2\theta}+{\cal E}(R).
\end{equation}
In the above Hamiltonian \ref{eqn:H0}, the
variables $(R, \theta, \phi)$ are the interatomic distance
between the two atoms, the polar angle of the dimer defined from the direction $\hat{\bf z}$, and
the azimuthal angle, respectively. $(P_R, P_\theta, P_\phi)$ are the corresponding
canonically conjugate momenta.
We assume that the polarization plane of the laser field is perpendicular to $\hat{\bf z}$, such that
its electric field is
\begin{equation}
\label{eq:EL}
{\bf E}(t)=\frac{F}{\sqrt{1+\epsilon^2}}\sqrt{f(t)}\left[ \hat{\bf x}\cos(\omega t+\phi)+
\hat{\bf y} \epsilon \sin(\omega t+\phi)\right],
\end{equation}
where $f(t)$ is the intensity envelope and $0 \le \epsilon \le 1$ is the ellipticity of the field.
The limit values $\epsilon=0$ and $\epsilon=1$ correspond to a linearly polarized laser field along the $\hat{\bf x}$
direction and to a circularly polarized laser field, respectively.
The envelope $f(t)$ is given by \cite{A182}
\begin{equation}
f(t)=\left\{ \begin{array}{ll} \displaystyle \sin^2 \left(\frac{\pi t}{2T_{\rm ru}}\right) & \mbox{if } 0\leq t<T_{\rm ru},\\ 
1 & \mbox{if } T_{\rm ru}\leq t<T_{\rm ru}+T_{\rm p},\\ 
 \displaystyle \sin^2\left(\frac{\pi (t-T_{\rm ru}-T_{\rm p}-T_{\rm rd})}{2T_{\rm rd}} \right) & \mbox{if } T_{\rm ru}+T_{\rm p}\leq t<T_{\rm ru}+T_{\rm p}+T_{\rm rd},\\
 0 & \mbox{elsewhere}. 
\end{array}\right.
\label{eq:pulse}
\end{equation}
where $T_{\rm ru}$, $T_{\rm p}$ and $T_{\rm rd}$ are the duration of the ramp-up, the plateau and the ramp-down
of the pulse, respectively. The field envelope \ref{eq:pulse} describes well experimental laser pulses \cite{A747}.
The interaction potential $V_{\rm int}$ between the molecule and the electric field of the laser pulse writes
\begin{equation}
\label{ham:inte1}
V_{\rm int}({\bf r})=-{\bf D}({\bf r})\cdot {\bf E}(t)-
\frac{1}{2}\ {\bf E}(t)\cdot \widehat  \alpha ({\bf r}) \cdot  {\bf E}(t),
\end{equation}
where ${\bf D}({\bf r})$ is the dipole moment function and $\widehat \alpha ({\bf r})$ is the polarizability tensor
of the dimer. In the nonresonant case \cite{A477,Seid97}, it is possible to average the dynamics
over the short temporal scale of the laser, i.e., $2\pi/\omega$,  such that we end up with the following
expression for the interaction potential~\ref{ham:inte1} :
\begin{equation}
\label{ham:inte2}
V_{\rm int}(R,\theta,\phi,t)=-f(t)\frac{F^2}{4(1+\epsilon^2)}\left[(\alpha_\parallel(R)-\alpha_\perp(R)) \sin^2\theta(\cos^2\phi+\epsilon^2\sin^2\phi)+(1+\epsilon^2) \alpha_\perp(R) \right].
\end{equation}
The functions $\alpha_{\parallel,\bot}(R)$ are, respectively, the parallel and the perpendicular
components of the molecular polarizability of the RbCs molecule \cite{polarizabilidad}.  
Thence, we write the total Hamiltonian $H$ of the system as the sum $H=H_0+V_{\rm int}$,
\begin{equation}
H(R,P_R,\theta,P_\theta,\phi, P_\phi,t)=\frac{P_R^2}{2\mu}+\frac{P_\theta^2}{2\mu R^2}+\frac{P_\phi^2}{2\mu R^2\sin^2\theta }+{\cal E}(R) +V_{\rm int}(R,\theta,\phi,t).
\label{eqn:Ham6D}
\end{equation}
\noindent
Hamiltonian~\ref{eqn:Ham6D} has 3+1$/$2 degrees of freedom
(the 1/2 degree of freedom is due to the explicit time-dependence through the envelope of the laser pulse). Moreover, Hamiltonian \ref{eqn:Ham6D}
presents the following invariant manifold ${\cal M}$
\begin{eqnarray}
\label{stable}
{\cal M} &=&\lbrace R,P_R, \theta=\pi/2,  P_{\theta}=0, \phi,
P_{\phi}\rbrace,
\end{eqnarray}
where the dynamics of the dimer is limited to planar motions confined to the polarization $xy$ plane of 
the laser field.
We reduce our study to the invariant manifold ${\cal M}$ such that, the degree of freedom associated with the motion outside the polarization plane is frozen, i.e., we consider $\theta=\pi/2$. The corresponding reduced Hamiltonian
in the manifold ${\cal M}$ has 2+1$/$2 degrees of freedom and it reads
\begin{equation}
H_{{\cal M}}(R,P_R,\phi, P_\phi,t)=\frac{P_R^2}{2\mu}+
\frac{P_\phi^2}{2\mu R^2 }+V_{{\cal M}}(R,\phi,t), 
\label{eqn:Ham4D}
\end{equation}
\noindent
where $V_{{\cal M}}$ is the total potential
energy surface on the manifold ${\cal M}$,
\begin{equation}
V_{{\cal M}}(R,\phi,t)={\cal E}(R)-f(t)\frac{F^2}{4(1+\epsilon^2)}\left[(\alpha_\parallel(R)-\alpha_\perp(R))
(\cos^2\phi+\epsilon^2\sin^2\phi)+(1+\epsilon^2) \alpha_\perp(R) \right]. 
\label{eqn:VM}
\end{equation}
\noindent
When the laser pulse is circularly polarized, $\epsilon=1$, the Hamiltonian
$H_{{\cal M}}$ has 1+1$/$2 degrees of freedom because the angle $\phi$ is cyclic
and the corresponding momentum $P_{\phi}$ is a constant of the motion.
In particular, during the plateau of the pulse, $f(t)=1$, and for $\epsilon=1$, the
system becomes integrable.
In what follows, we consider electric fields of intensity between 0 and $7\times 10^{12}~{\rm W}\cdot{\rm cm}^{-2}$, which
roughly correspond to electric field strengths up to $F \approx 10^{-2}$ a.u. 
\begin{figure}
\centerline{\includegraphics[scale=.35]{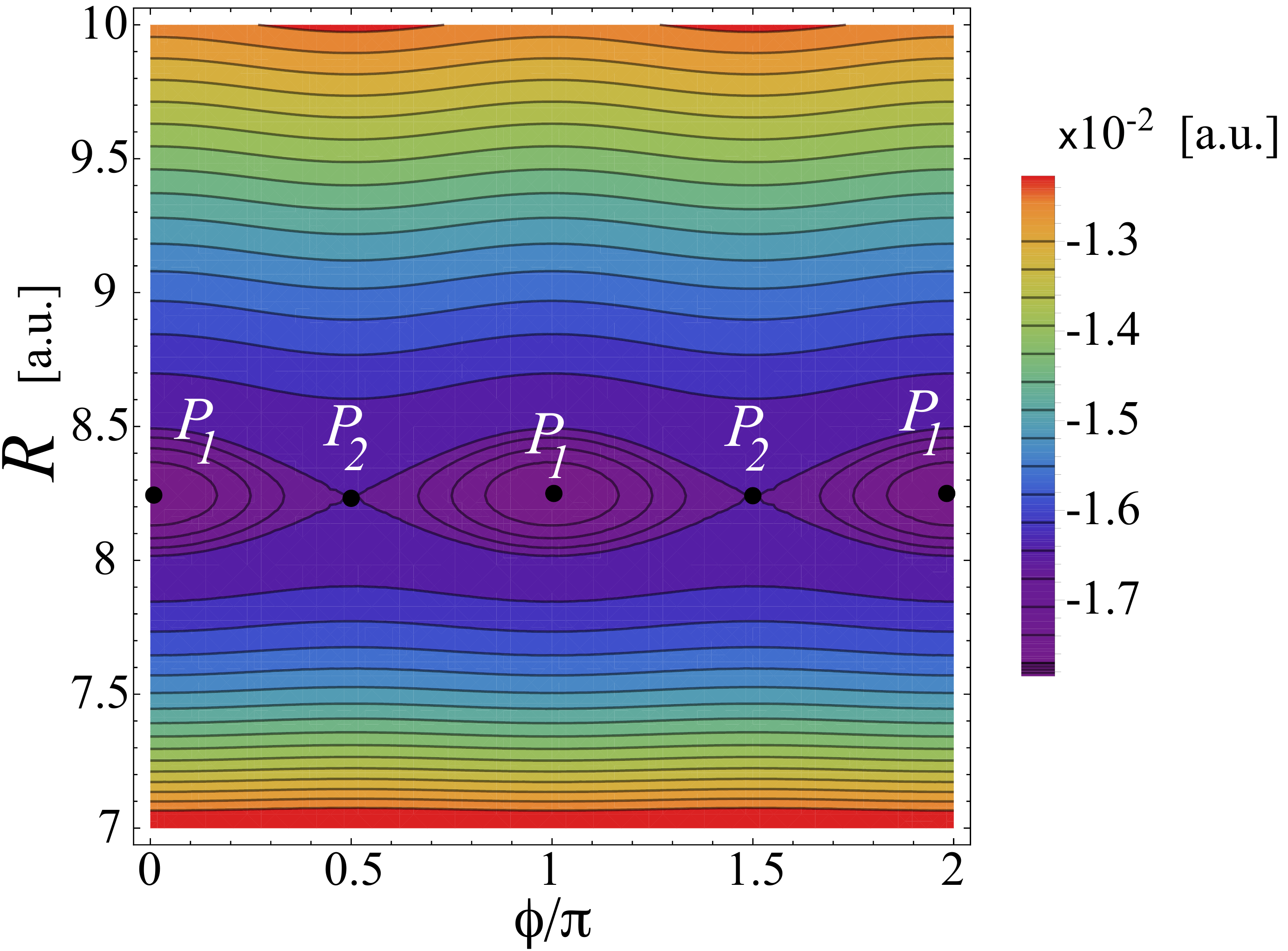}}

\centerline{ \includegraphics[scale=.35]{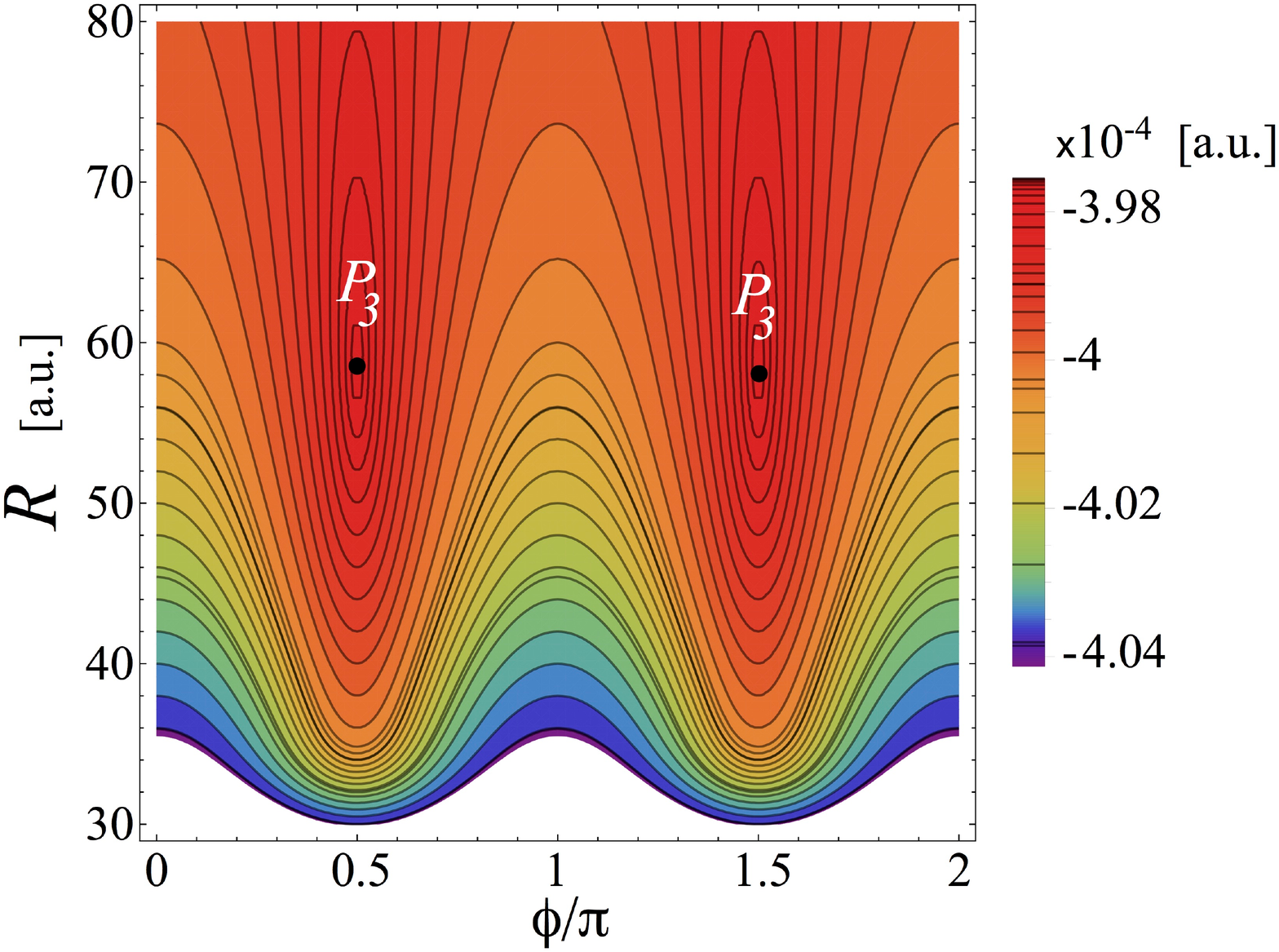}}
\caption{Equipotential curves of the potential energy
surface $V_{\cal M}(R,\phi,t)$  during the plateau (i.e., with $f(t)=1$)
for a laser field strength $F=1.5 \times 10^{-3}$ a.u. and ellipticity $\epsilon=0.5$.}
\label{fi:potencial}
\end{figure}
It is practical for classical calculations to have an analytical representation for
the potential energy surface $V_{{\cal M}}(R,\phi,t)$. Therefore, we have fitted the available data for ${\cal E}(R)$ \cite{pec} and
$\alpha_{\parallel,\bot}(R)$  \cite{polarizabilidad}
to three appropriate functional forms. The fitting of ${\cal E}(R)$ includes its long-range
behavior described by
\begin{equation}
\label{pecLR}
{\cal E}_{LR}(R) = -\frac{b_6}{R^6} - \frac{b_8}{R^8} - \frac{b_{10}}{R^{10}}.
\end{equation}
For the $^1\Sigma^+$ RbCs,  the $b_i$ coefficients in Eq.~\ref{pecLR} can be found in
Ref.~\cite{marinescu}.
In the medium and long range, the behavior of the polarizabilities $\alpha_{\parallel,\bot}(R)$
is well described by the Silberstein expressions
\cite{Silberstein,A495}
\begin{eqnarray}
\label{Silberstein}
\alpha_\parallel^{LR}(R) &=& \frac{\alpha_{\rm RbCs} + 4 \alpha_{\rm Rb} \alpha_{\rm Cs}/R^3}
{1 - 4 \alpha_{\rm Rb} \alpha_{\rm Cs}/R^6},\nonumber\\
& & \\
\alpha_\bot^{LR}(R) &=& \frac{\alpha_{\rm RbCs} - 2 \alpha_{\rm Rb} \alpha_{\rm Cs}/R^3}
{1 -  \alpha_{\rm Rb} \alpha_{\rm Cs}/R^6}\nonumber,
\end{eqnarray}
\noindent
where $\alpha_{\rm Rb}\approx313$ a.u. and $\alpha_{\rm Cs}\approx394$ a.u.
are the atomic polarizabilities of the two species and $\alpha_{\rm RbCs}=\alpha_{\rm Rb} + \alpha_{\rm Cs}$.
However, in the short range, the Silberstein expressions \ref{Silberstein} do not provide a correct 
description of the polarizabilities $\alpha_{\parallel,\bot}(R)$ because of the divergences. In analytical calculations,
instead of using Eq.~\ref{Silberstein} for the long-range behavior, we use the asymptotic
limits of Eq.~\ref{Silberstein} given by
\begin{eqnarray}
\label{Silberstein2}
\alpha_\parallel^{LR}(R) &\approx& \alpha_{\rm RbCs} + \frac{4 \alpha_{\rm Rb} \alpha_{\rm Cs}}{R^3},\nonumber\\
& & \\
\alpha_\bot^{LR}(R) &\approx& \alpha_{\rm RbCs} - \frac{2 \alpha_{\rm Rb} \alpha_{\rm Cs}}{R^3}.\nonumber
\end{eqnarray}

\subsection{Analysis of the potential energy surface}
\begin{figure}
\includegraphics[width=0.8\textwidth]{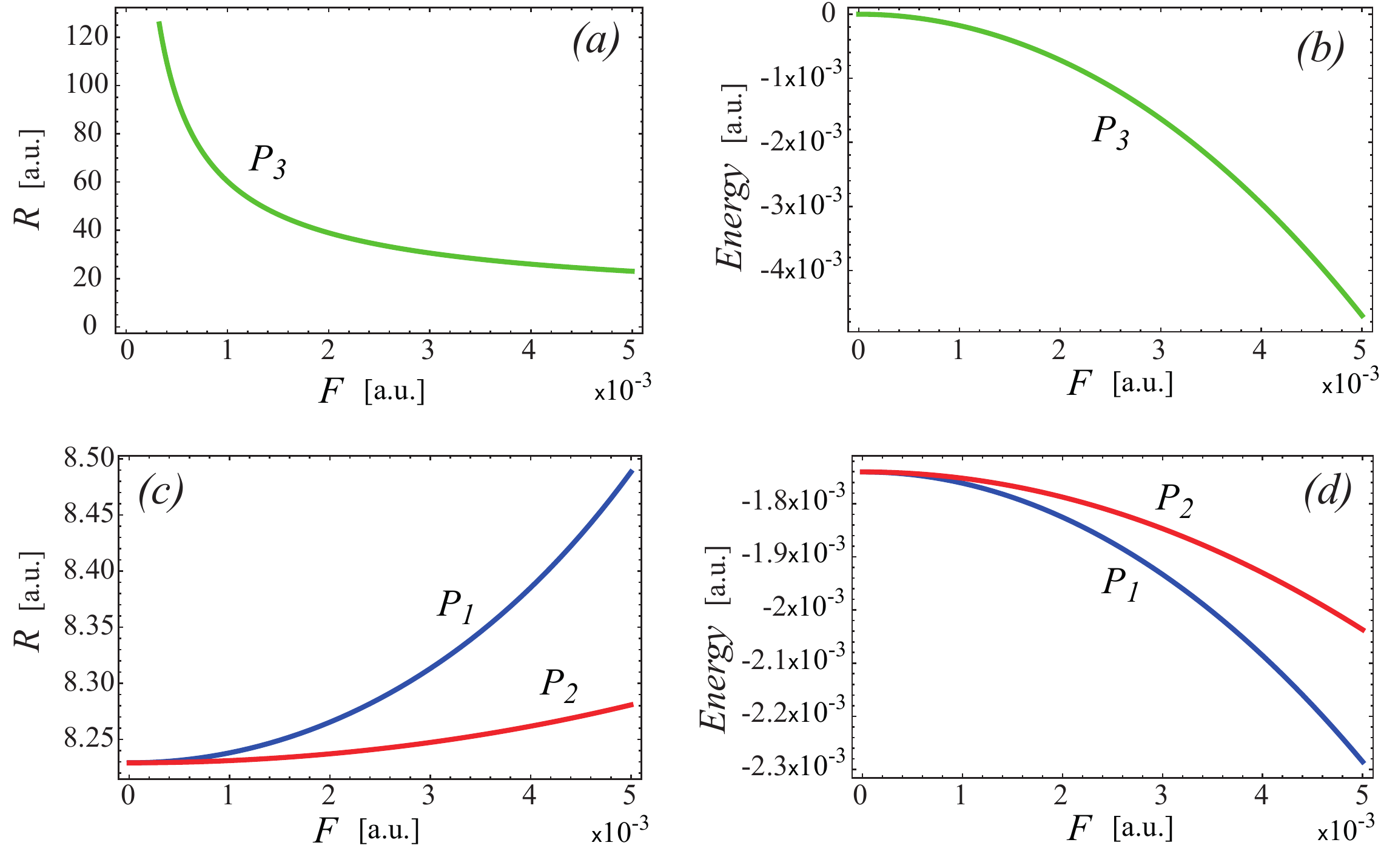}
 \caption{Evolution of the position and energy of the maxima $P_3$ [upper panels (a) and (b)], the
 saddle points $P_2$ [blue lines in the upper panels (a) and (b)] and the minima $P_1$
  [red lines in the upper panels (a) and (b)] of the potential energy surface
 $V_{{\cal M}}(R,\phi,t)$ during the plateau (i.e., with $f(t)=1$) as a function of the amplitude of the electric field $F$.
All figures for the same ellipticity value $\epsilon=0.5$.
 }
 \label{fi:figure02} 
 \end{figure}
\begin{figure}
\includegraphics[width=0.8\textwidth]{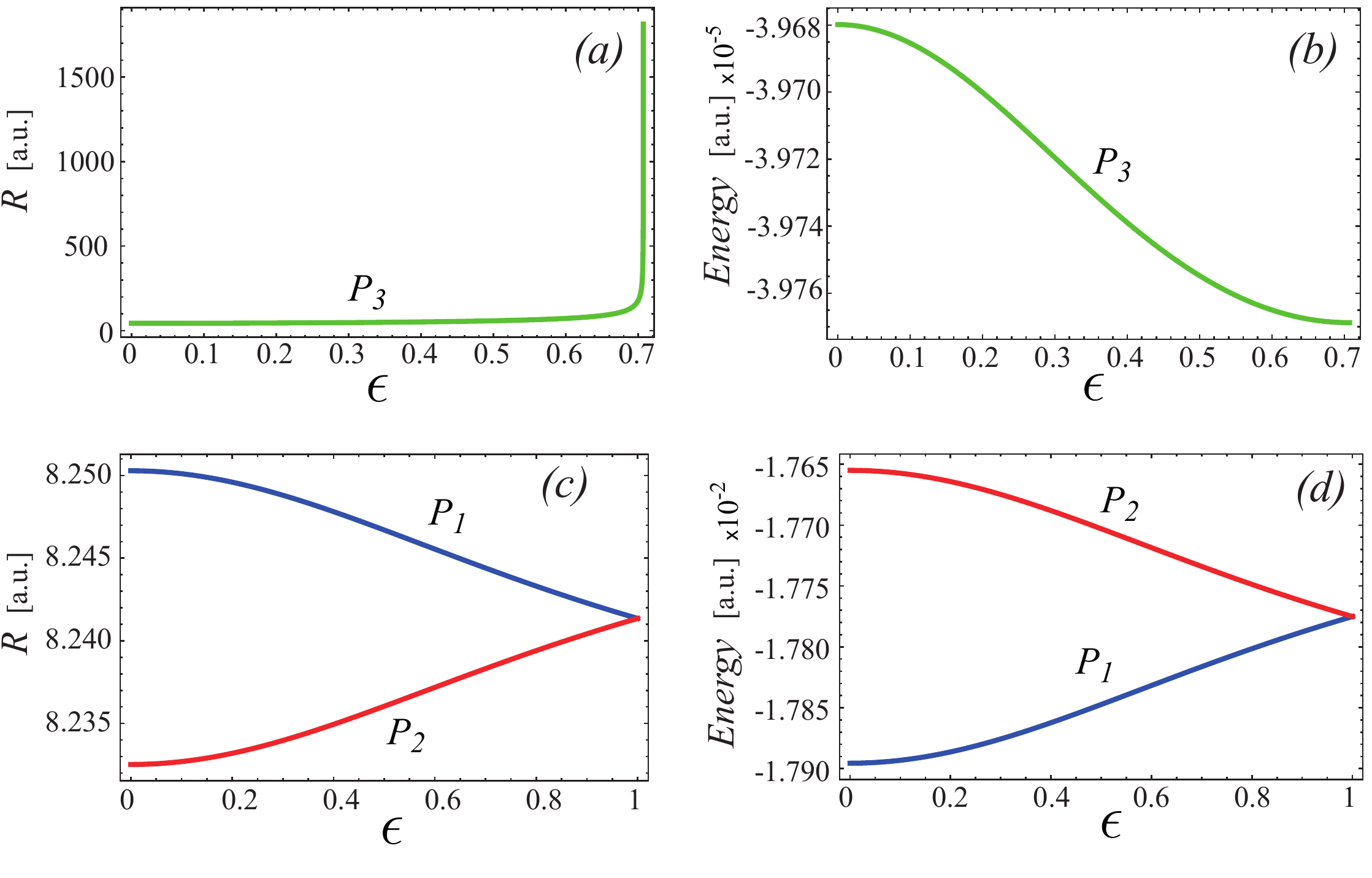}
 \caption{Evolution of the position and energy of the maxima $P_3$ [upper panels (a) and (b)], the
 saddle points $P_2$ [blue lines in the upper panels (a) and (b)] and the minima $P_1$
  [red lines in the upper panels (a) and (b)] of the potential energy surface
 $V_{{\cal M}}(R,\phi,t)$ during the plateau ($f(t)=1$) as a function of the ellipticity $\epsilon$.
All figures for the same electric field value $F=1.5\times 10^{-3}$ a.u.}
 \label{fi:figure03} 
 \end{figure}

The landscape of the potential energy surface $V_{{\cal M}}(R,\phi,t)$ during the plateau
(i.e., where $f(t)=1$) is mainly determined by its critical points. We notice that these critical points are fixed points of the dynamics if $P_R=P_\phi=0$. It is straightforward to see that they
appear along the directions $\phi=0$ and $\pi/2$~$(\mbox{mod } \pi)$, as it is depicted in
Fig.~\ref{fi:potencial}, for $F=1.5\times 10^{-3}$ a.u. In the short-range domain, there are two minima $P_1$ at $\phi=0$ and $\pi$, and two saddle points $P_2$ at
$\phi=\pi/2$ and $3\pi/2$. In the long-range domain, there are two maxima $P_3$ at $\phi=\pi/2$ and $3\pi/2$. Thence, when the energy of the system
is below the energy of the saddle points $P_2$, the (classical) dimer is confined to move into one of the potential wells created by the minima $P_1$. However, when the energy of the system is above the energy of the saddle points
$P_2$, the molecule can describe complete rotations. Due to the energy hills
around the maxima $P_3$ created by the polarizability, the largest values of the
intermolecular distance $R$ are obtained along the directions $\phi=0$, $\pi/2$ and $\pi$.
Obviously, the location and energy of the critical points depend on the values of the parameters $F$ and $\epsilon$.
In this way, as we can observe in Fig.~\ref{fi:figure02}(a)-(b), as the electric field strength $F$ increases, the
maxima $P_3$ approaches the saddle points $P_2$
and their energy decreases. In the same way, as $F$ increases, the depth of the potential wells determined by
$P_1$ and $P_2$ increases, while their position shows a slight
increase [see Fig.~\ref{fi:figure02}(c)-(d)].
The influence of the parameter $\epsilon$ is shown in Fig.~\ref{fi:figure03}:
As ellipticity is increased, the maximum $P_3$ quickly moves off the saddle points $P_2$ and its energy decreases. At a critical ellipticity $\epsilon \gtrsim 1/\sqrt{2}$, this maximum disappears [see Fig.~\ref{fi:figure03}(a)-(b)]. On the other side, as the ellipticity
$\epsilon$ increases, the depth of the potential wells determined by $P_1$ and $P_2$ decreases, such that for
$\epsilon=1$, they come into coincidence and they disappear.
In this way, in the integrable case of $\epsilon=1$, the
equipotential curves of $V_{{\cal M}}$ are straight lines of constant $R$ value. This is the
expected landscape for a potential energy surface which only depends
on the interatomic distance $R$.

The described evolution of the critical points can be seen analytically by looking at the short-range behavior near the bottom of the well created by the potential ${\cal E}(R)$: We assume that in this region the effect of the interaction with the laser field is a small perturbation of the unperturbed Hamiltonian, we obtain the following expressions for the positions of $P_1$ and $P_2$:
\begin{eqnarray*}
&& R(P_1)\approx R_{\rm min}+\frac{F^2}{4(1+\epsilon^2){\cal E}^{''}(R_{\rm min})}\left[ \alpha^\prime_\parallel(R_{\rm min})+\epsilon^2 \alpha^\prime_\perp(R_{\rm min}) \right],\\
&& R(P_2)\approx R_{\rm min}+\frac{F^2}{4(1+\epsilon^2){\cal E}^{''}(R_{\rm min})}\left[ \alpha^\prime_\perp(R_{\rm min})+\epsilon^2 \alpha^\prime_\parallel(R_{\rm min}) \right],
\end{eqnarray*}
where $\alpha^\prime_{\parallel,\perp}=d \alpha_{\parallel,\perp}/dR$ and
$R_{\rm min}$ is the location of the minimum of the potential ${\cal E}(R)$. From these expressions, we notice that these positions increases like $F^2$ as $F$ increases, with a higher increase for the position of $P_1$ since $\alpha^\prime_\parallel(R_{\rm min})$ is larger than $\alpha^\prime_\perp(R_{\rm min})$. In addition, we notice that the position of $P_1$ (respectively $P_2$) decreases (respectively increases) with increasing ellipticity, and the positions of $P_1$ and $P_2$ coincide for $\epsilon=1$. Actually, when $\epsilon=1$ all the points with $R=R(P_1)=R(P_2)$ are fixed points, irrespective of the value of the angle $\phi$, due to symmetry.  The energy of these points are given by
\begin{eqnarray*}
&& V_{\cal M}(P_1)\approx {\cal E}(R_{\rm min})-\frac{F^2}{4(1+\epsilon^2)}\left[ \alpha_\parallel(R_{\rm min})+\epsilon^2 \alpha_\perp(R_{\rm min}) \right],\\
&& V_{\cal M}(P_2)\approx {\cal E}(R_{\rm min})-\frac{F^2}{4(1+\epsilon^2)}\left[ \alpha_\perp(R_{\rm min})+\epsilon^2 \alpha_\parallel(R_{\rm min}) \right].
\end{eqnarray*}
These energies decrease like $F^2$ as $F$ increases, and $V_{\cal M}(P_1)$ decreases faster than $V_{\cal M}(P_2)$ since $\alpha_\parallel(R_{\rm min})>\alpha_\perp(R_{\rm min})$. From these expressions, one can show that the energy of $P_2$ (respectively $P_1$) decreases (respectively increases) when $\epsilon$ increases. As expected, their values are equal when $\epsilon=1$. 

Using the long-range expressions for the potentials, an approximate expression of the location of $P_3$ can be derived
$$
R(P_3)\approx \left(\frac{4b_6(1+\epsilon^2)}{F^2(1-2\epsilon^2)\alpha_{\rm Rb}\alpha_{\rm Cs}} \right)^{1/3},
$$
from which we clearly see the singularity at $\epsilon=1/\sqrt{2}$ and its decrease as $F^{-2/3}$ as $F$ increases. The associated energy varies as
$$
V_{\cal M}(P_3)\approx -\frac{F^2}{4}\alpha_{\rm RbCs}+\frac{F^4(1-2\epsilon^2)\alpha_{\rm Rb}^2\alpha_{\rm Cs}^2}{16b_6(1+\epsilon^2)^2}. 
$$ 
Given the value of the parameters for RbCs, the second term is much smaller than the first one. Therefore the energy decreases like $-F^{2}$ as $F$ increases, and there is a very weak dependence of the energy with respect to the ellipticities in the domain where $P_3$ exists. 
In the limit $R \longrightarrow \infty$, the potential curve
${\cal E}(R)$ tends to zero, while $\alpha_\parallel(R)$ and $\alpha_\perp (R)$ tend to $\alpha_{\rm RbCs}$. Using
the potential energy surface $V_{{\cal M}}(R,\phi,t)$ during the plateau ($f(t)=1$) 
the dissociation threshold $E_d$ is thus given by
\begin{equation}
E_d \approx -\frac{F^2}{4} \alpha_{\rm RbCs} .
\label{dissociation}
\end{equation}
It is worth noticing that the dissociation threshold does not depend on the ellipticity $\epsilon$.

\section{Phase Space Structure}
\label{Phase}
\begin{figure}
\centerline{\includegraphics[scale=.6]{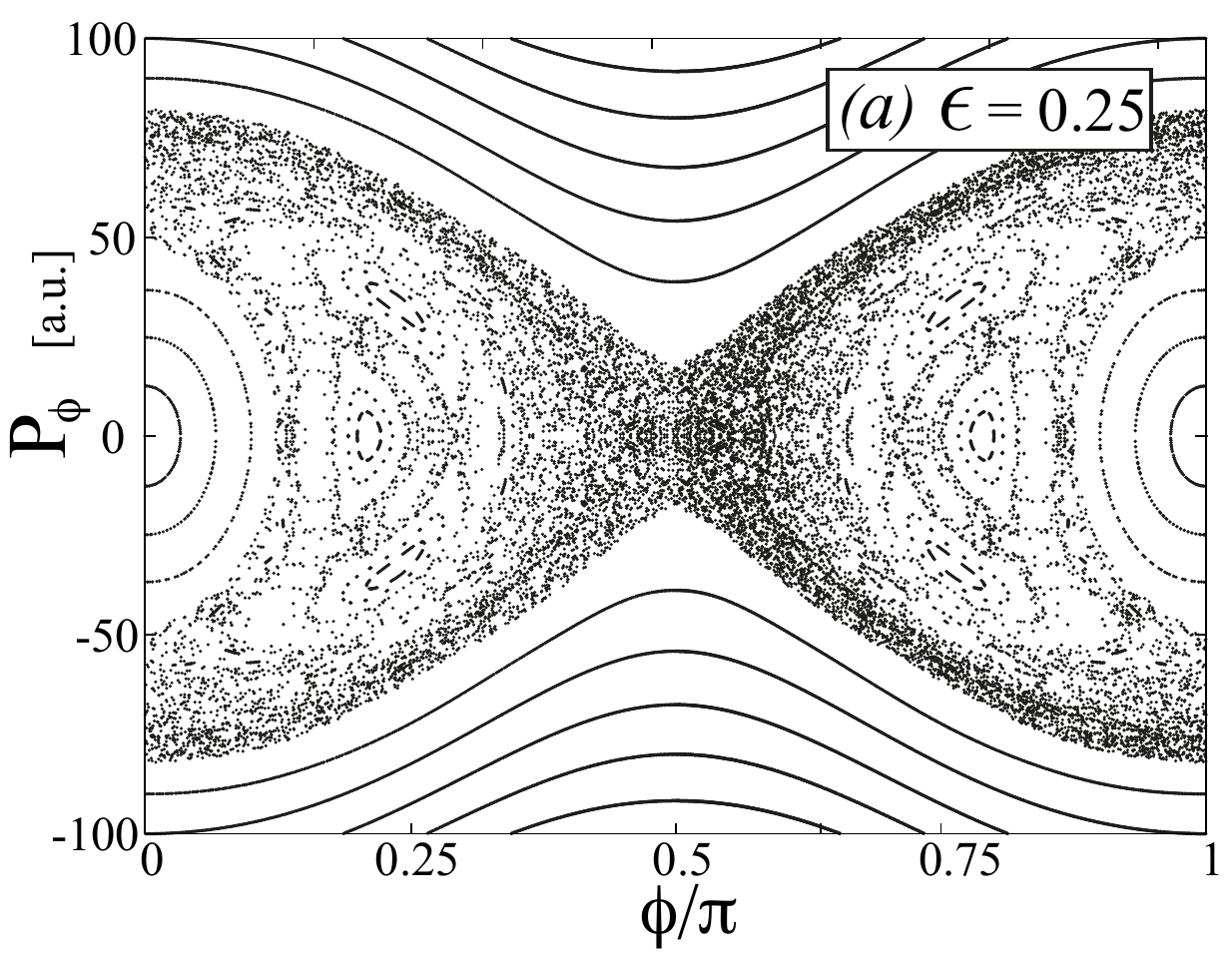}\quad \includegraphics[scale=.6]{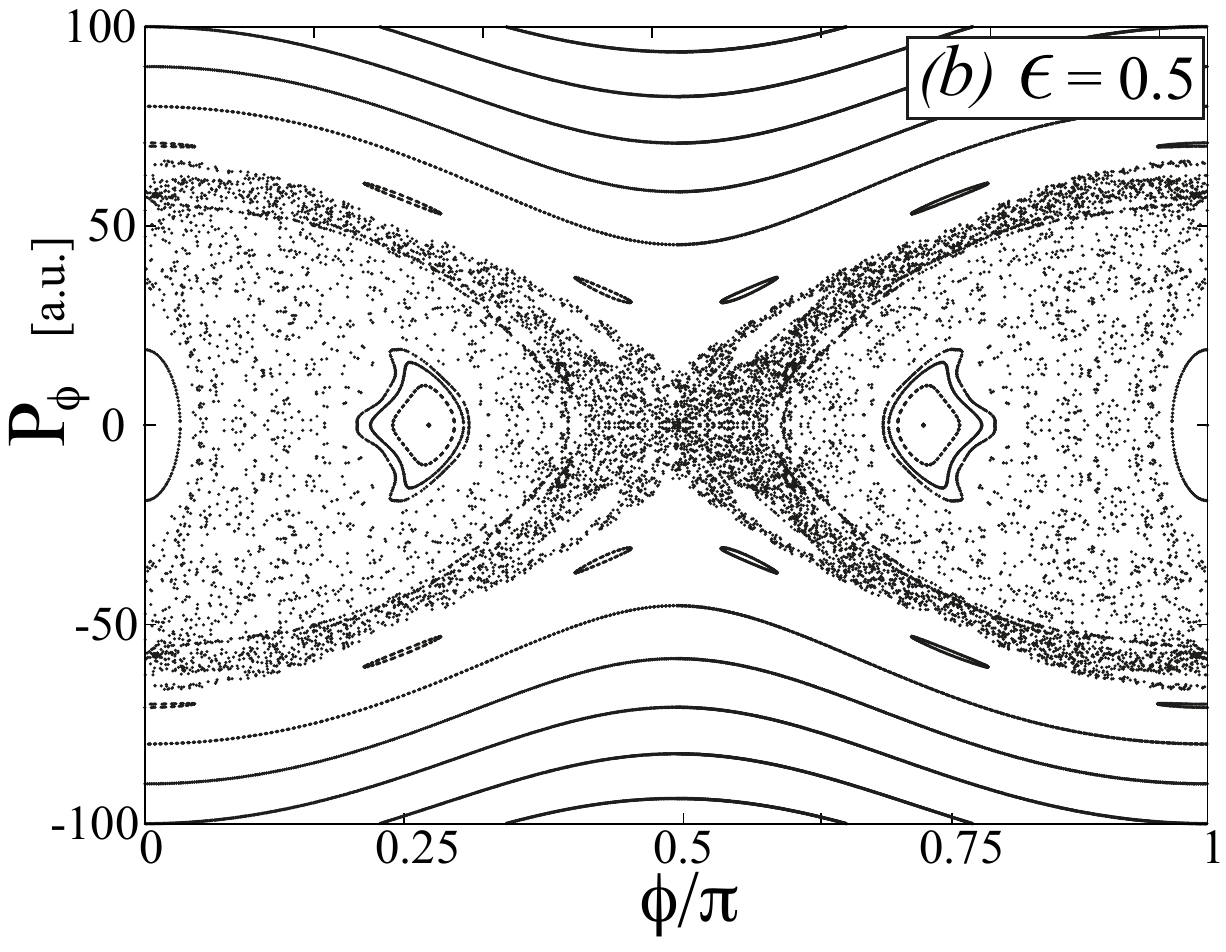}}

\centerline{\includegraphics[scale=.6]{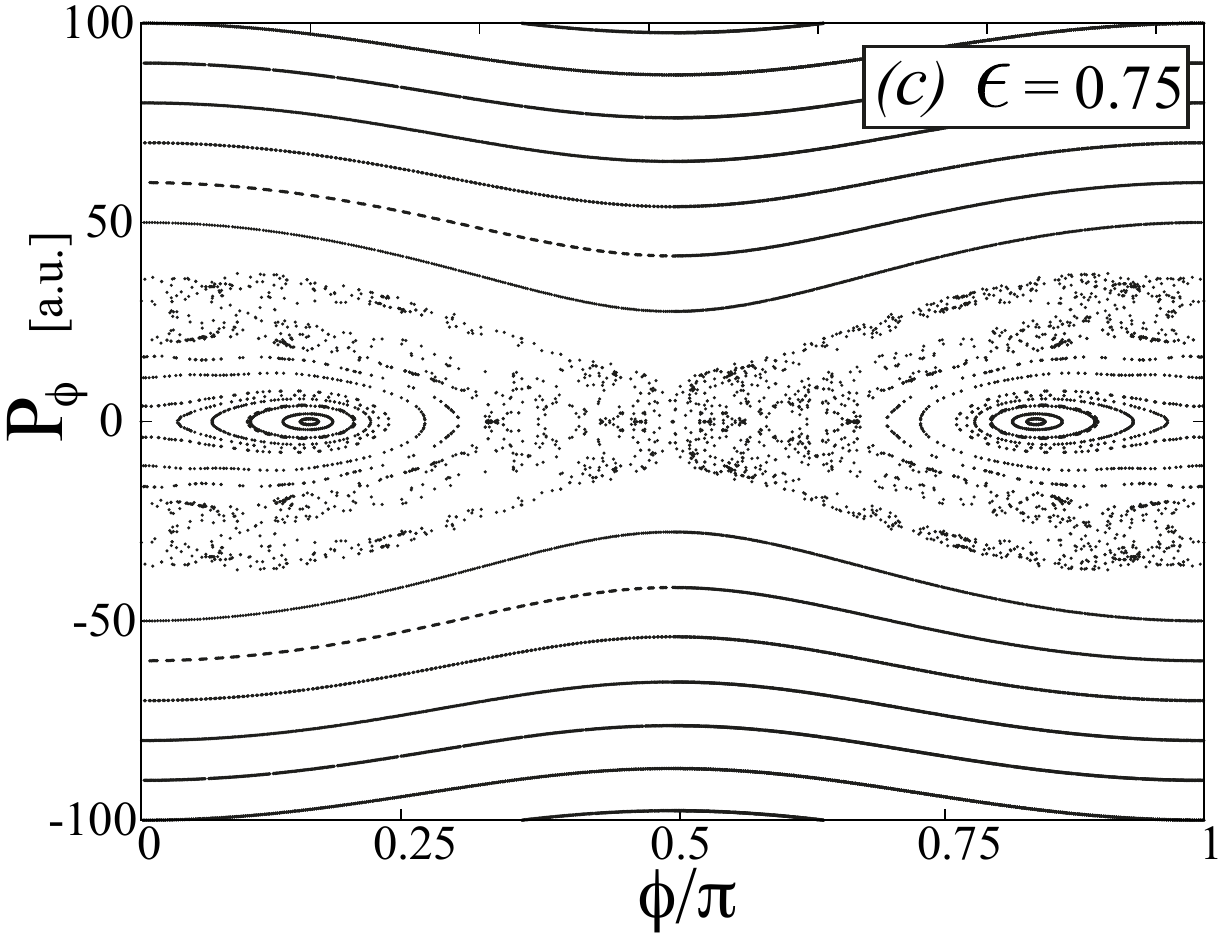}\quad \includegraphics[scale=.6]{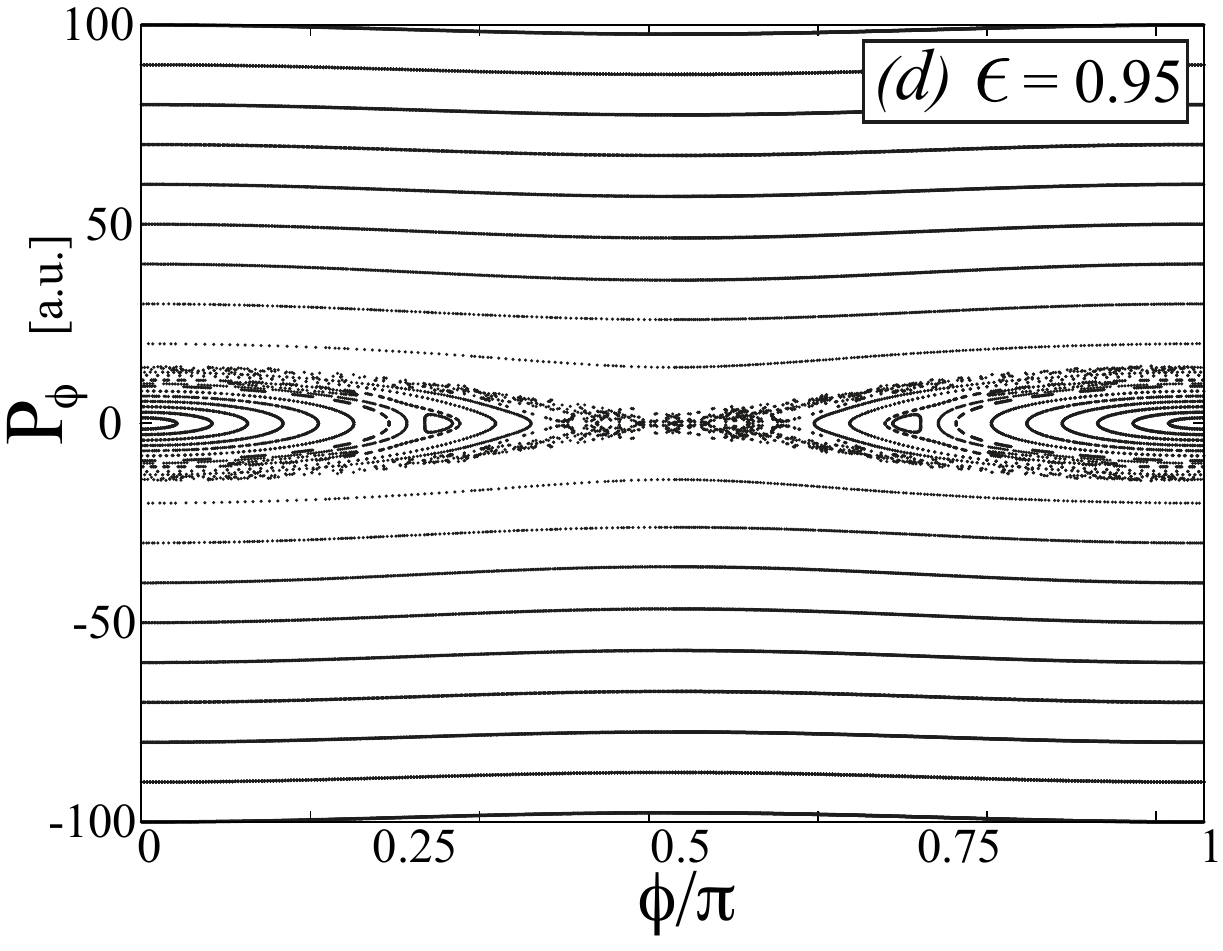}}
\caption{Poincar\'e sections ($P_R=0, \dot P_R>0$) of Hamiltonian \ref{eqn:Ham4D}
with $F=2 \times 10^{-3}$ a.u., an energy $E=-7.08\times10^{-4}$ a.u.\
and for $\epsilon=0.25$, 0.5, 0.75 and 0.95.}
\label{fi:sos1}
\end{figure}

Because during the plateau Hamiltonian \ref{eqn:Ham4D} has two degrees
of freedom, Poincar\'e sections constitute a very convenient tool for visualizing its phase space
structures. In order to get information from the orbits populating the surfaces of section, we consider
the long-term dynamics of the trajectories such that they are calculated
for a time much larger than the duration
of the pulse, typically up to $2\times 10^4$ ps.
A convenient Poincar\'e map is $P_R=0$ with $\dot P_R>0$, such that the
trajectories are mapped onto the plane $(\phi, P_{\phi})$.
In particular, we are interested in investigating the changes of these Poincar\'e sections
as the external
parameters $F$ and $\epsilon$ are varied, and when
the energy $E$ of the dimer is near the dissociation threshold given by Eq.~\ref{dissociation}.
For $F=2\times10^{-3}$ a.u., the dissociation energy is given by $E_d\approx -7.07\times10^{-4}$ a.u.
For an energy of $E=-7.08\times10^{-4}$ a.u., i.e., slightly lower than $E_d$,
and for $\epsilon=0.25$, 0.5, 0.75 and 0.95, the corresponding Poincar\'e sections are shown in Fig.~\ref{fi:sos1}.
In the four sections depicted in Fig.~\ref{fi:sos1} we find the same generic structure that
resembles a chaotic pendulum. Indeed, we find regular rotational and vibrational trajectories and a
rotational chaotic layer around the hyperbolic fixed point at $\phi=\pi/2$. There also
appears a chaotic region of
vibrational orbits that seems to be disconnected from the rotational chaotic region.
As expected, for
increasing ellipticity, the size of the vibrational region as well as the width of the chaotic layer decrease
because, as the ellipticity tends to unity, $\epsilon \rightarrow 1$, the system approaches its
aforementioned integrable limit where the angle $\phi$ is cyclic (and where the momentum $P_{\phi}$ is conserved), such that 
the phase space is only populated with rotational orbits.

\begin{figure}
\centerline{\includegraphics[scale=.6]{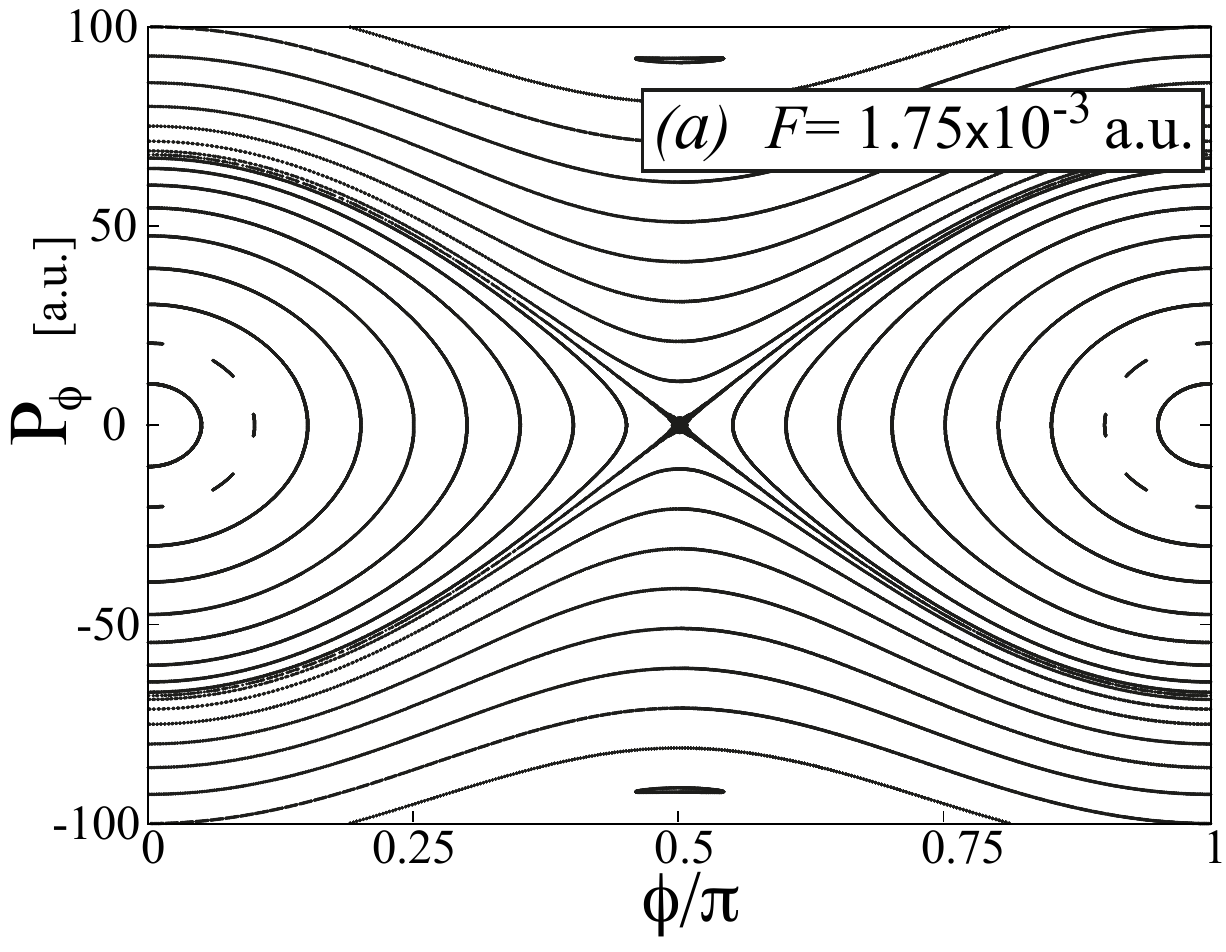}\quad \includegraphics[scale=.6]{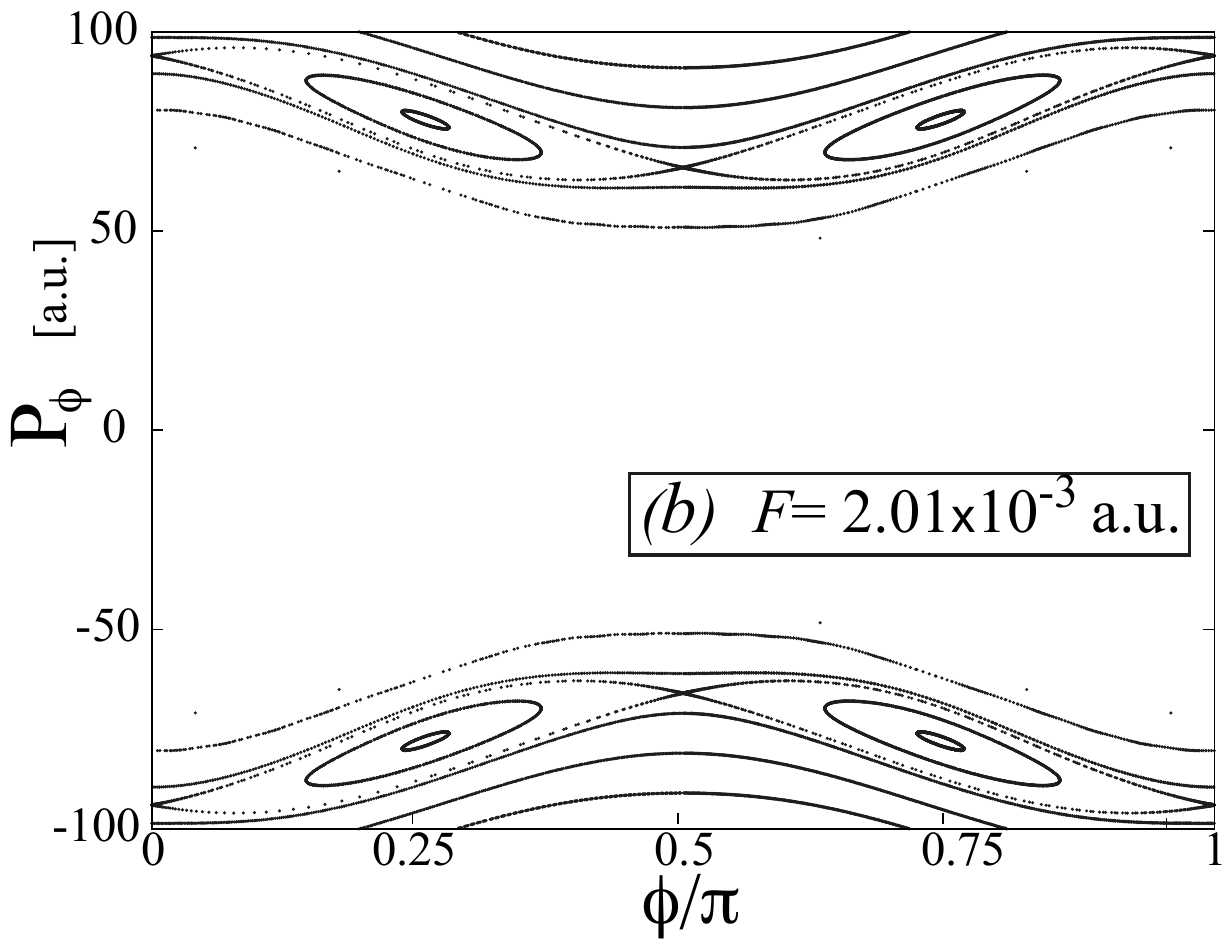}}
\caption{Poincar\'e sections ($P_R=0, \dot P_R>0$) of Hamiltonian \ref{eqn:Ham4D}
 for $F=1.75 \times 10^{-3}$ a.u. and $F=2.01 \times 10^{-3}$ a.u. Both sections have been computed
 with ellipticity $\epsilon=0.25$  and energy $E=-7.08\times10^{-4}$ a.u.}
\label{fi:sos2}
\end{figure}

Now, starting from the configuration depicted in the Poincar\'e map of Fig.~\ref{fi:sos1}(b)
for $\epsilon=0.5$, $E=-7.08\times10^{-4}$ a.u. and $F=2 \times 10^{-3}$ a.u., we
change the electric field strength in order to study its effect on the phase space structure.
When $F$ is slightly smaller, e.g. $F=1.75 \times 10^{-3}$ a.u., the phase portrait undergoes significant changes. Except for a tiny chaotic region around the unstable fixed point located
at $(\pi, 0)$, the Poincar\'e map of Fig.~\ref{fi:sos2}(a) for $F=1.75 \times 10^{-3}$ a.u.
is equivalent to that of a pendulum, with 
a phase space structure made of regular rotational and vibrational orbits.
The reason of this quick change from a mixed regular-chaotic behavior to a fairly regular
behavior is that the dissociation threshold $E_d$ given by Eq.~\ref{dissociation}
quadratically increases with $F$. Then, even an slight decrease of $F$ leaves the system
 well below its dissociation threshold, which in general moves nonlinear systems to more regular
 behaviors.
When the electric field increases from the starting value $F=2 \times 10^{-3}$ a.u., the
dissociation threshold $E_d$ given by Eq.~\ref{dissociation} decreases and 
its effect on the dynamics is even more dramatic because
most of the chaotic trajectories in the
Poincar\'e map of Fig.~\ref{fi:sos1}(b) rapidly become dissociation orbits.
For example, for the slightly larger value
$F=2.01 \times 10^{-3}$ a.u., we have that $E_d\approx -7.14\times10^{-4}$ a.u., and the
Poincar\'e section of Fig.~\ref{fi:sos2}(b) presents a large central empty region which
corresponds to dissociation trajectories. In the situation depicted in Fig.~\ref{fi:sos2}(b), only the
rotational orbits with $P_{\phi}$ value large enough remain isolated from the dissociation channels along the directions $\phi=\pi/2$ and $3\pi/2$.

In order to interpret the basic structures behind these Poincar\'e sections, we build an effective model to measure the size of the resonant island where the chaotic motion is confined to. The assumption is a short-range one and it is based on the fact that the values of $R$ stay close to the minimum of the potential well. The effective Hamiltonian becomes
$$
H_{\rm eff}=\frac{P_\phi^2}{2\mu R_{\rm min}^2}-\frac{F^2(1-\epsilon^2)}{8(1+\epsilon^2)}\left(\alpha_\parallel(R_{\rm min})-\alpha_\perp(R_{\rm min}) \right)\cos 2\phi. 
$$ 
It is the Hamiltonian of a pendulum with a stable equilibrium at $\phi=0$ and an unstable equilibrium at $\phi=\pi/2$ (mod $\pi$).  
The width of the resonant island is then given by
$$
\Delta P_\phi=\sqrt{2} R_{\rm min}F\sqrt{\mu \left(\alpha_\parallel(R_{\rm min})-\alpha_\perp(R_{\rm min}) \right)}\sqrt{\frac{1-\epsilon^2}{1+\epsilon^2}}.
$$
It increases linearly with $F$ and decreases as $\epsilon$ approaches 1.  Inside the separatrix, the degree of freedom $(R,P_R)$ cannot be frozen and there is a complex interaction between the two degrees of freedom leading to chaotic behaviors. It should be noted that inside the separatrix, typical trajectories experience large excursions away from the equilibrium points, so the full potential is needed to describe the dynamics. 
\begin{figure}
\centerline{\includegraphics[scale=.8]{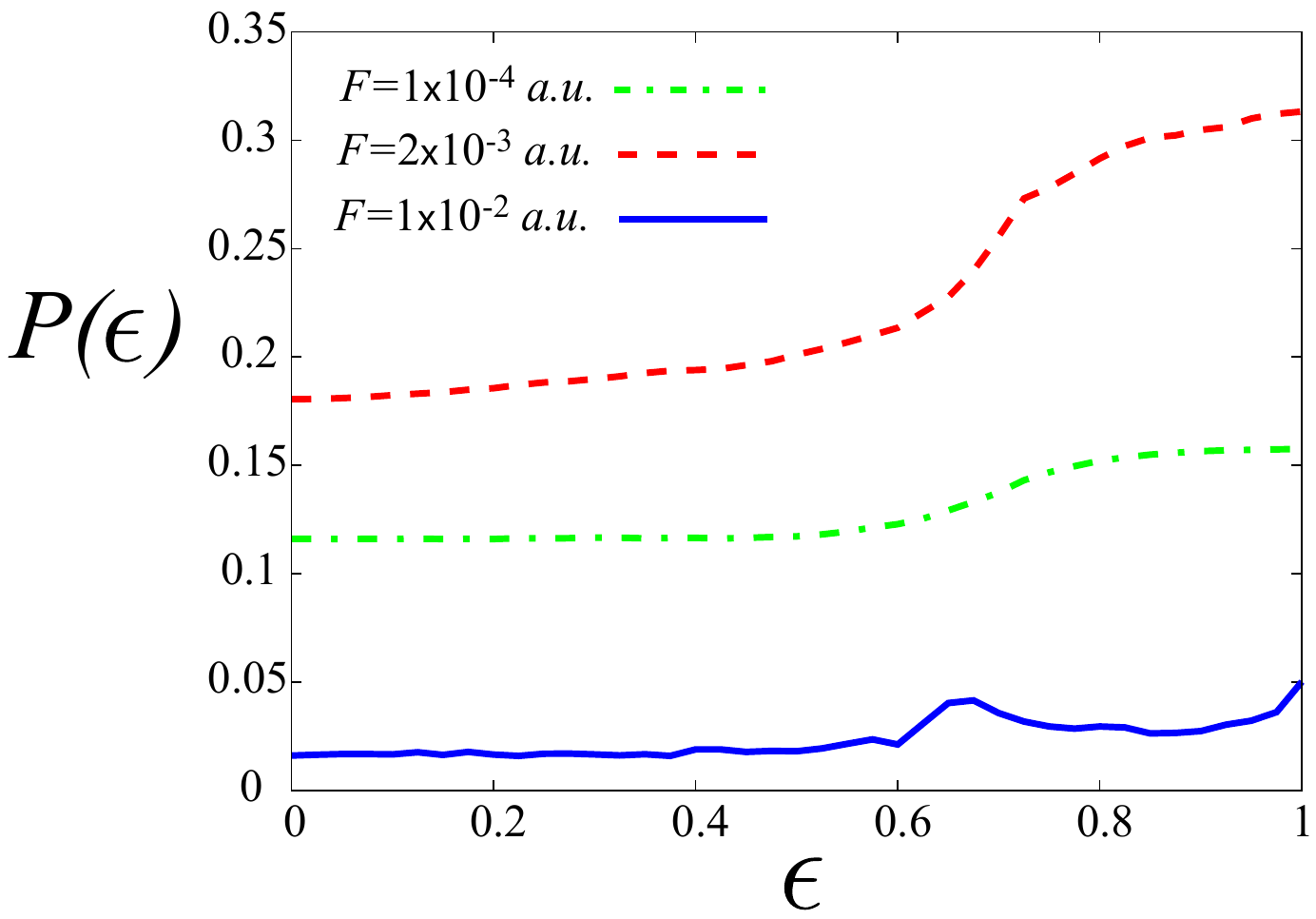}}
\caption{Formation probabilities obtained as a function of the ellipticity $\epsilon$
for  $F=1 \times 10^{-4}$ a.u. (dashed-dotted green line), $F=2 \times 10^{-3}$ a.u.
(shaded red line) and $F= 1\times10^{-2}$ a.u. (solid blue line).
The probabilities have been calculated for an initial ensemble of
initial conditions with energy ${E}_0=3\times 10^{-9}$.
The parameters of the pulse are $T_{\rm ru}=15~\mbox{ps}$, $T_{\rm p}=70~\mbox{ps}$ and $T_{\rm rd}=15~\mbox{ps}$}
\label{fi:probaR200}
\end{figure}

\section{Formation Probability as a Function of the Ellipticity}
\label{driving}
\subsection{Numerical results}
In this section, we study the influence of the elliptically polarized laser pulse \ref{eq:EL} on the
creation of bound RbCs molecular states. To this end, for each ellipticity between 0 (linear polarization)
and 1 (circular polarization) and for different values of the field strength $F$, we take a
large ensemble of free pairs of Rb-Cs atoms.
At $t=0$, all the initial conditions $(R_0, P_R^0, \phi_0, P_{\phi}^0)$
of the ensemble have the same energy
$E_0=3\times 10^{-9}$ a.u., which roughly
corresponds to $T=1$ mK, the typical temperature of a sample of cold
atoms in current photoassociation experiments \cite{A859,A331}.
In our calculations, the initial values of $P_\phi^0$ 
are taken to be zero, while the values of $\phi_0$ are chosen randomly in $[0,2\pi]$.
The initial interatomic distances $R_0$
are chosen in the interval $[R_{\rm m}, R_{\rm M}]=[6.2329, 200]$ a.u., where
$R_{\rm m}$ is the inner turning point of the phase trajectory given by the  ``free" Hamiltonian
\begin{equation}
E \equiv H_{{\cal M}}^0=\frac{P_R^2}{2\mu}+
\frac{P_\phi^2}{2\mu R^2 }+{\cal E}(R),
\label{eqn:Ham4DFree}
\end{equation}
\noindent
for $E=E_0=3\times 10^{-9}$ a.u. and $P_{\phi}=0$. Finally, the values of $P_R^0$ are given by
\[
P_R^0=\pm \sqrt{2 \mu (E_0 - {\cal E}(R_0))}.
\]
\noindent
Then, we propagate the ensemble of trajectories by integrating numerically the equations of motion obtained from
Hamiltonian~\ref{eqn:Ham4D} up to the pulse duration $t_{\rm final}= T_{\rm ru} + T_{\rm p}+T_{\rm rd}$.
For $t \ge t_{\rm final}$, the energy of each trajectory of the ensemble remains constant and the dynamics
is governed by the free Hamiltonian~\ref{eqn:Ham4DFree}.
In order to have a bound trajectory at the end of
the pulse, the final energy $E_{\rm final}$ of a given trajectory
at $t_{\rm final}$ has to be lower than the height of the potential barrier
of Hamiltonian~\ref{eqn:Ham4DFree}, and the final intermolecular
distance has to be smaller than the location of the barrier. In order to get an approximate formation criterion, we
use the long--range behavior for  the potential energy:
\[
{\cal E}(R)\approx -\frac{b_6}{R^6}.
\]
The effective potential in Eq.~\ref{eqn:Ham4DFree} is given by
\[
{\cal E}_{\rm eff}(R)=\frac{P_\phi^2}{2\mu R^2 }-\frac{b_6}{R^6}.
\]
The above effective potential has a maximum for $R=R_*$,
\[
R_*=\left(\frac{ 6 \mu b_6}{P_\phi^2}\right)^{1/4},
\]
and the height of the corresponding potential barrier is
\[
{\cal E}_*=\frac{2 b_6}{R_*^6}.
\]
Therefore, at the end of the pulse, formation occurs for
$(R(t_{\rm final}), P_R(t_{\rm final}), \phi(t_{\rm final}),P_\phi(t_{\rm final}))$ such that
\begin{equation}
\label{criterium}
E_{\rm final}=\frac{P_R^2(t_{\rm final})}{2\mu}+
\frac{P_\phi^2(t_{\rm final})}{2\mu R(t_{\rm final})^2 }+{\cal E}(R(t_{\rm final})) \le \frac{2 b_6}{R_*^6},\quad \mbox{and}
\quad R(t_{\rm final}) \le R_*.
\end{equation}
Once the ensemble has been propagated up to $t=t_{\rm final}$, we consider the proportion
of formed RbCs dimers, i.e., the formation probability.
For the total duration of the pulse we use $t_{\rm final}=100$ ps, a value easily achieved in current
experiments \cite{A747}. For the chosen values of the parameters, the values for the formation probabilities are approximately the same if a rough negative energy criterion is used, i.e., $E_{\rm final}<0$ for formation. 

In Fig.~\ref{fi:probaR200}, we show the the formation
probability $P(\epsilon)$ as a function of the ellipticity $\epsilon$
for a laser profile $T_{\rm ru}=T_{\rm rd}=15$ ps and $T_{\rm p}=70$ ps. We consider three different regimes of laser amplitudes: low, intermediate and high laser amplitudes.
Increasing laser amplitude, we move from one regime to the next. Here typically, we choose $F=1\times10^{-4}$ a.u. (low amplitude), $F=2\times10^{-3}$ a.u. (intermediate amplitude) and $F=1\times10^{-2}$ a.u. (high amplitude). 
The low, intermediate and high laser amplitudes display different behaviors. For low and intermediate amplitudes, the formation probability varies very weakly with increasing ellipticity, then rather abruptly increases around $\epsilon\approx 0.7$.
At high amplitudes, the formation probability shows a different behavior: $P(\epsilon)$
is very small for laser ellipcities below $\epsilon  \lesssim 0.7$ and then it presents a peak of formation
around that value $\epsilon\approx 0.7$, such that, for larger values of $\epsilon$, the
formation probability $P(\epsilon)$ almost saturates. It is worth noticing that the location of this peak does
not change significantly by changing the duration of the pulse. As we will show in the next section, the
behavior of the formation probability, and in particular its significant increase around
$\epsilon\approx 0.7$, can be explained as only a function of the long--range
parameters of the polarizabilities of the dimer and the ellipticity $\epsilon$ of the laser.

\subsection{Results from a static approximation}

\begin{figure}
\includegraphics[width=0.6\textwidth]{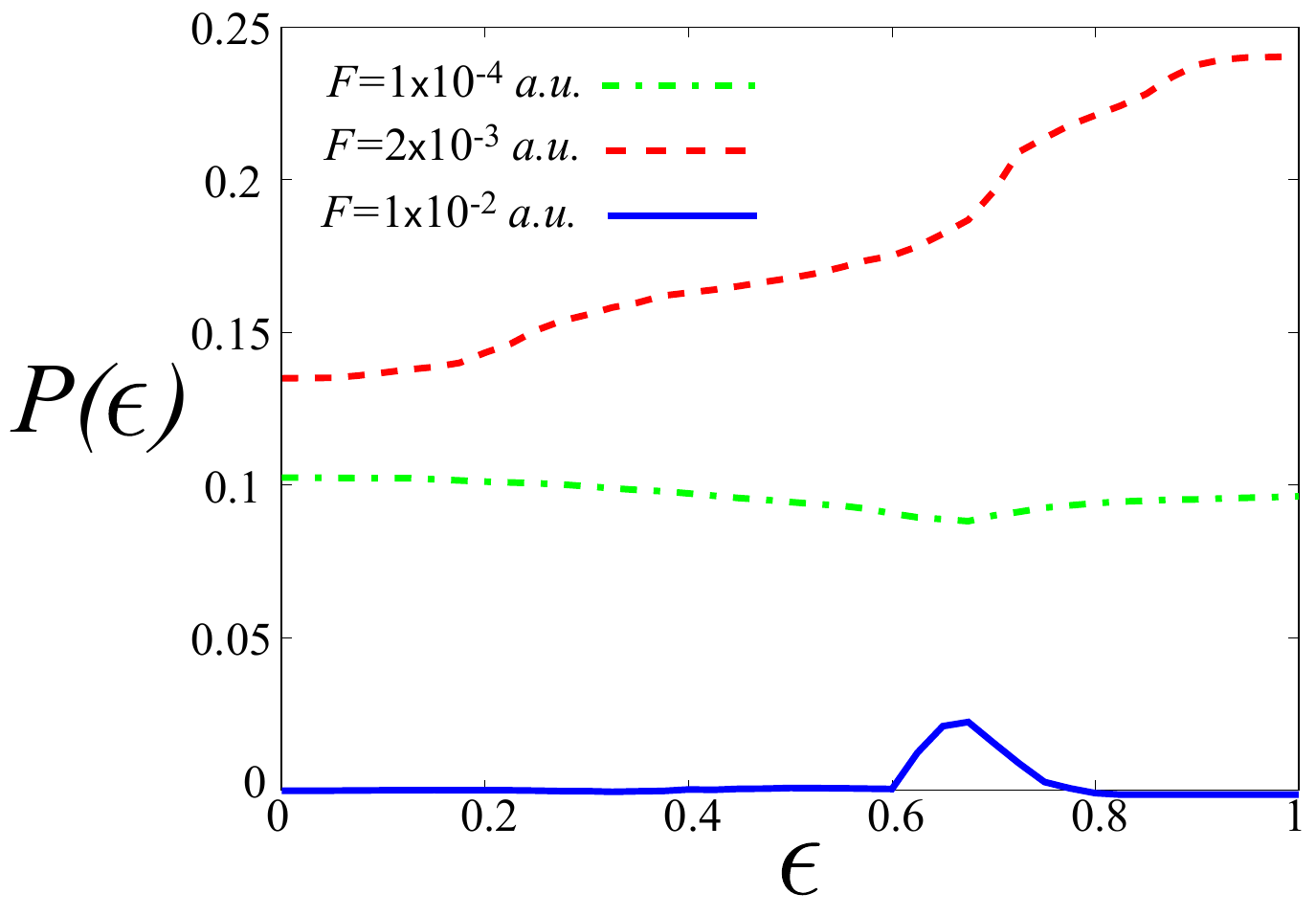}
 \caption{\label{fig:ProbaS4Dell} Formation probability obtained with Eq.~\ref{eqn:Ef} as a function of ellipticity $\epsilon$ for $F=1 \times 10^{-4}$ a.u. (dashed-dotted green line), $F=2 \times 10^{-3}$ a.u. (shaded red line)
and $F= 1\times10^{-2}$ a.u. (solid blue line).
The probabilities have been calculated for an initial ensemble of
initial conditions with energy ${ E}_0=3\times 10^{-9}$.
The parameters of the pulse are $T_{\rm ru}=15~\mbox{ps}$, $T_{\rm p}=70~\mbox{ps}$ and $T_{\rm rd}=15~\mbox{ps}$.
}%
 \end{figure}

\begin{figure}
\includegraphics[scale=.6]{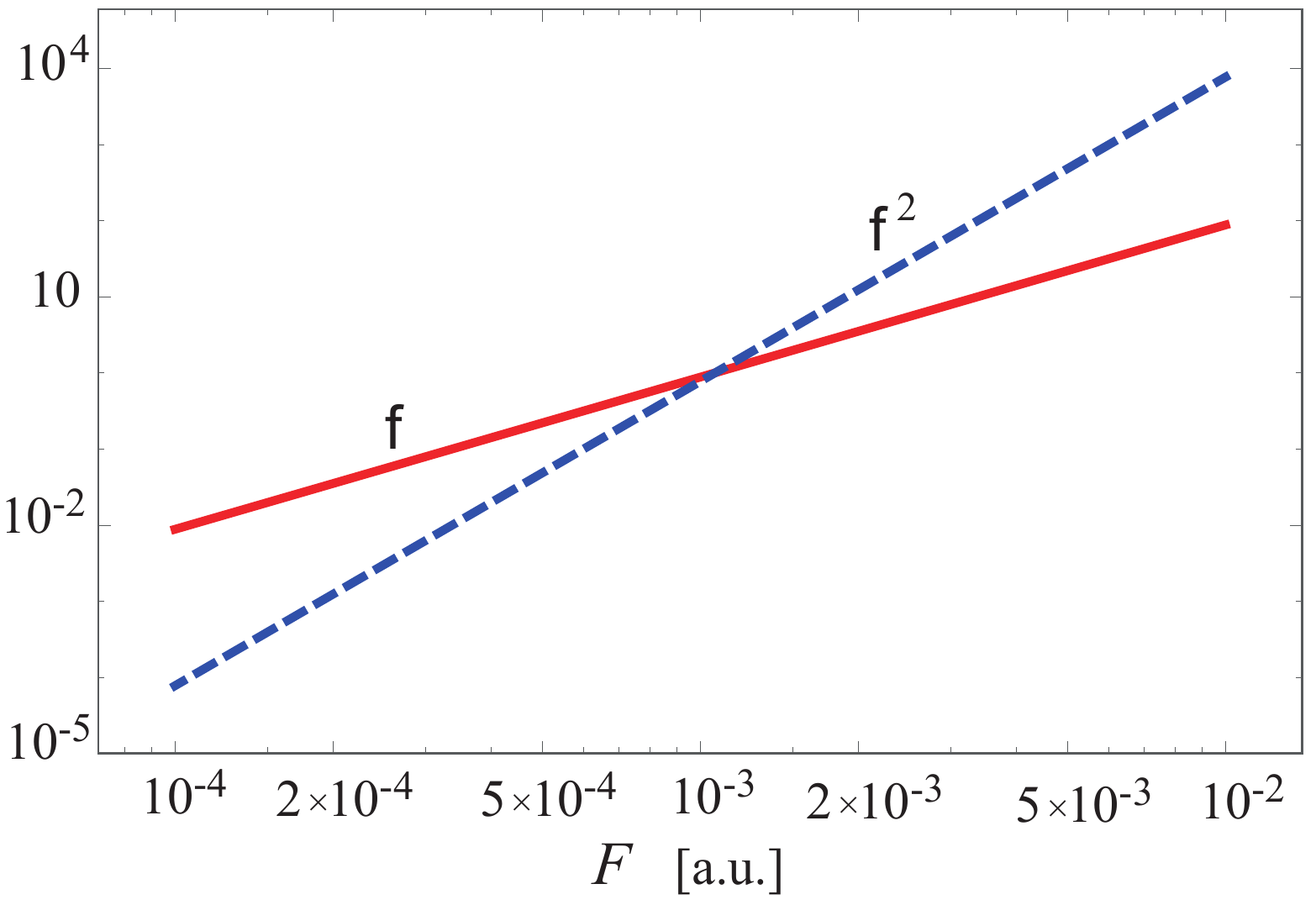}
 \caption{\label{fig:paraF} Evolution of the parameter ${\rm f}$ (red solid line) and ${\rm f}^2$ (blue dashed line)
 as a function of the field strength $F$ in
 interval $F\in [10^{-4}, 10^{-2}]$. The parameters of the pulse are $T_{\rm ru}=15~\mbox{ps}$, $T_{\rm p}=70~\mbox{ps}$ and $T_{\rm rd}=15~\mbox{ps}$, such that ${\rm f} = {\rm f}^2$ at $F\approx 10^{-3}$ a.u.}
 \end{figure}

We assume that the pulse is sufficiently short and the mass sufficiently large, so that the atoms
have no time to move. Nonetheless, they acquire a momentum shift induced by the laser
pulse, and this is sufficient to ensure dimer formation for selected initial conditions. In order to be a bit more quantitative, we consider the four-dimensional case with Hamiltonian~\ref{eqn:Ham4D}. A similar reasoning can be done for
the six-dimensional case with Hamiltonian~\ref{eqn:Ham6D}. 
The inverse of the reduced mass is the small parameter. The spatial coordinates are given by
\begin{eqnarray*}
&& R = R_0+O(\mu^{-1}),\\
&& \phi = \phi_0 +O(\mu^{-1}). 
\end{eqnarray*}
The equations of motion for the momenta arising from Hamiltonian~\ref{eqn:Ham4D} can be written as:
\begin{eqnarray}
\label{eq:movi1}
&& \dot{P_R}= \frac{P_\phi^2}{\mu R^3}-{\cal E}^\prime (R)+f(t)\frac{F^2}{4}{\cal F}(R,\phi;\epsilon),\\
\label{eq:movi2}
&& \dot{P_\phi} = f(t) \frac{F^2}{4} R {\cal G}(R,\phi;\epsilon),
\end{eqnarray}
where ${\cal E}^\prime (R)=d {\cal E}(R)/dR$, and the functions $\cal F$ and $\cal G$ are given by
\begin{eqnarray*}
&& {\cal F}(R,\phi;\epsilon)=\frac{(\cos^2\phi+\epsilon^2\sin^2\phi)}{1+\epsilon^2}(\alpha^\prime_\parallel(R)-\alpha^\prime_\perp(R))+\alpha^\prime_\perp(R),\\
&& {\cal G}(R,\phi;\epsilon)=-\frac{\alpha_\parallel(R)-\alpha_\perp(R)}{R}\frac{1-\epsilon^2}{1+\epsilon^2} 
\sin 2\phi.
\end{eqnarray*}
The integration of Eqs.~\ref{eq:movi1}-\ref{eq:movi2}
for the duration of the pulse, up to order $O(\mu^{-1})$, leads to 
\begin{eqnarray}
\label{eq:integration1}
&& P_R=P_R^0+\Delta P_R +O(\mu^{-1}),\\
&& P_\phi =\Delta P_\phi +O(\mu^{-1}),\label{eq:integration2}
\end{eqnarray}
since $P_\phi^0=0$, and where we assume that ${\cal E}^\prime (R) \approx 0$. Then, the
laser induced momentum shifts are given by:
\begin{eqnarray}
&& \Delta P_R=\frac{F^2(T_{\rm ru}+2T_{\rm p}+T_{\rm rd})}{8}{\cal F}(R_0,\phi_0;\epsilon),\label{eq:transfer1}\\
\label{eq:transfer2}
&& \Delta P_\phi=\frac{F^2(T_{\rm ru}+2T_{\rm p}+T_{\rm rd})}{8} R_0 {\cal G}(R_0,\phi_0;\epsilon),
\end{eqnarray}
where we have used the fact that 
$$
\int_0^{T_{\rm ru}+T_{\rm p}+T_{\rm rd}} f(t){\rm d}t= \frac{T_{\rm ru}}{2}+T_{\rm p}+\frac{T_{\rm rd}}{2}.
$$
We observe that the parameters of the laser field $F$,
$T_{\rm ru}$, $T_{\rm p}$ and $T_{\rm rd}$ are involved in the momentum transfer quantities~\ref{eq:transfer1}-\ref{eq:transfer2} with a single parameter ${\rm f}$ of the form
\[
{\rm f} =\frac{F^2(T_{\rm ru}+2T_{\rm p}+T_{\rm rd})}{8}.
\]
Using Eqs.~\ref{eq:integration1}-\ref{eq:transfer2} and the free Hamiltonian~\ref{eqn:Ham4DFree}, the final energy is approximately given by
\begin{equation}
\label{eqn:Ef}
{E}_f={ E}_0+\frac{1}{\mu}P_R^0 \Delta P_R+\frac{1}{2\mu}(\Delta P_R)^2
+\frac{1}{2\mu R_0^2}(\Delta P_\phi)^2.
\end{equation}
All the terms in the above equation are of order $1/\mu$ and the neglected terms are of order $1/\mu^2$. We use this equation to determine an approximate formation probability. We consider a set of initial conditions in the same way as in Fig.~\ref{fi:probaR200} and we look at the subset which
holds the criterion~\ref{criterium} or an approximate negative energy criterion. In Fig.~\ref{fig:ProbaS4Dell}, we represent the resulting formation probabilities as a function of ellipticity for laser intensities $F=1 \times 10^{-4}$ a.u., $F=2 \times 10^{-3}$ a.u.
and $F= 1\times10^{-2}$ a.u.
We notice that for intermediate and large field intensities $F=2 \times 10^{-3}$ a.u.
and $F= 1\times10^{-2}$ a.u., they display the same behavior as in Fig.~\ref{fi:probaR200}, with a pronounced increasing
of formation around $\epsilon\approx 0.7$. However, for low field strength $F=1 \times 10^{-4}$ a.u., the
approximate expression \ref{eqn:Ef} does not provide good results. We will go back to this disagreement
for low field intensities. At this point, we analyze the different regimes displayed by the formation probability curves in the long-range approximation using the following formulas: 
\begin{eqnarray*}
&& {\cal F}(R,\phi;\epsilon)=-\frac{3\alpha_{\rm Rb}\alpha_{\rm Cs}}{R^4}\left(1+3\frac{1-\epsilon^2}{1+\epsilon^2}\cos 2\phi \right),\\
&& {\cal G}(R,\phi;\epsilon)=-\frac{6\alpha_{\rm Rb}\alpha_{\rm Cs}}{R^4}\frac{1-\epsilon^2}{1+\epsilon^2}\sin 2\phi, 
\end{eqnarray*}
obtained using Eq.~\ref{Silberstein2}.
A good approximation for the formation probability criterium is that the final energy is negative. Therefore formation occurs for initial conditions $(R_0,\phi_0)$ satisfying
\begin{equation}
E_f = {E}_0- 3 {\rm f} P_R^0 \frac{\alpha_{\rm Rb}\alpha_{\rm Cs}}{\mu R_0^4} g_1(\phi_0,\epsilon)+9 {\rm f}^2\frac{\alpha_{\rm Rb}^2\alpha_{\rm Cs}^2 }{2\mu R_0^8} g_2(\phi_0,\epsilon)< 0, \label{eqn:EfLR}
\end{equation}
where the functions $g_{1,2}(\phi,\epsilon)$ are
\begin{eqnarray}
\label{eq:g1}
g_1(\phi_0,\epsilon)&=&  1+3\frac{1-\epsilon^2}{1+\epsilon^2} \cos 2\phi_0,\\[2ex]
\label{eq:g2}
g_2(\phi_0,\epsilon)&=&\left( 1+3\frac{1-\epsilon^2}{1+\epsilon^2} \cos 2\phi_0 \right)^2+4 \left(\frac{1-\epsilon^2}{1+\epsilon^2} \right)^2 \sin^2 2\phi_0.
\end{eqnarray}
In Eq.~\ref{eqn:EfLR}, there are three terms: one independent of $\rm f$, one linear in $\rm f$ and one proportional to ${\rm f}^2$. The linear term is the only one which can be negative, so its contribution is essential for formation. The subtle balance between these three terms explains at least qualitatively the different behaviors observed in the formation probability curves as the parameters are varied.
Depending on the value of the field strength $F$, there are basically three regimes for most values of the ellipticity: one where $\rm f$ is so small that the ${\rm f}^2$ term can be neglected, one where the quadratic term in ${\rm f}$ is of the same order as the linear term, and one where the quadratic term is mostly dominant. These three regimes are clearly identified  in
Fig.~\ref{fig:paraF} where ${\rm f}$ and ${\rm f}^2$ are plotted as a function of $F$.
Furthermore, these regimes are expected to be very sensitive to the value of the field strength because
${\rm f}$ and ${\rm f}^2$ are quadratic and quartic functions in $F$, with $F\ll 1$.
We analyze below the formation probability in these three regimes.

\subsubsection{For large values of $\rm f$} 
\begin{figure}
\includegraphics[width=0.6\textwidth]{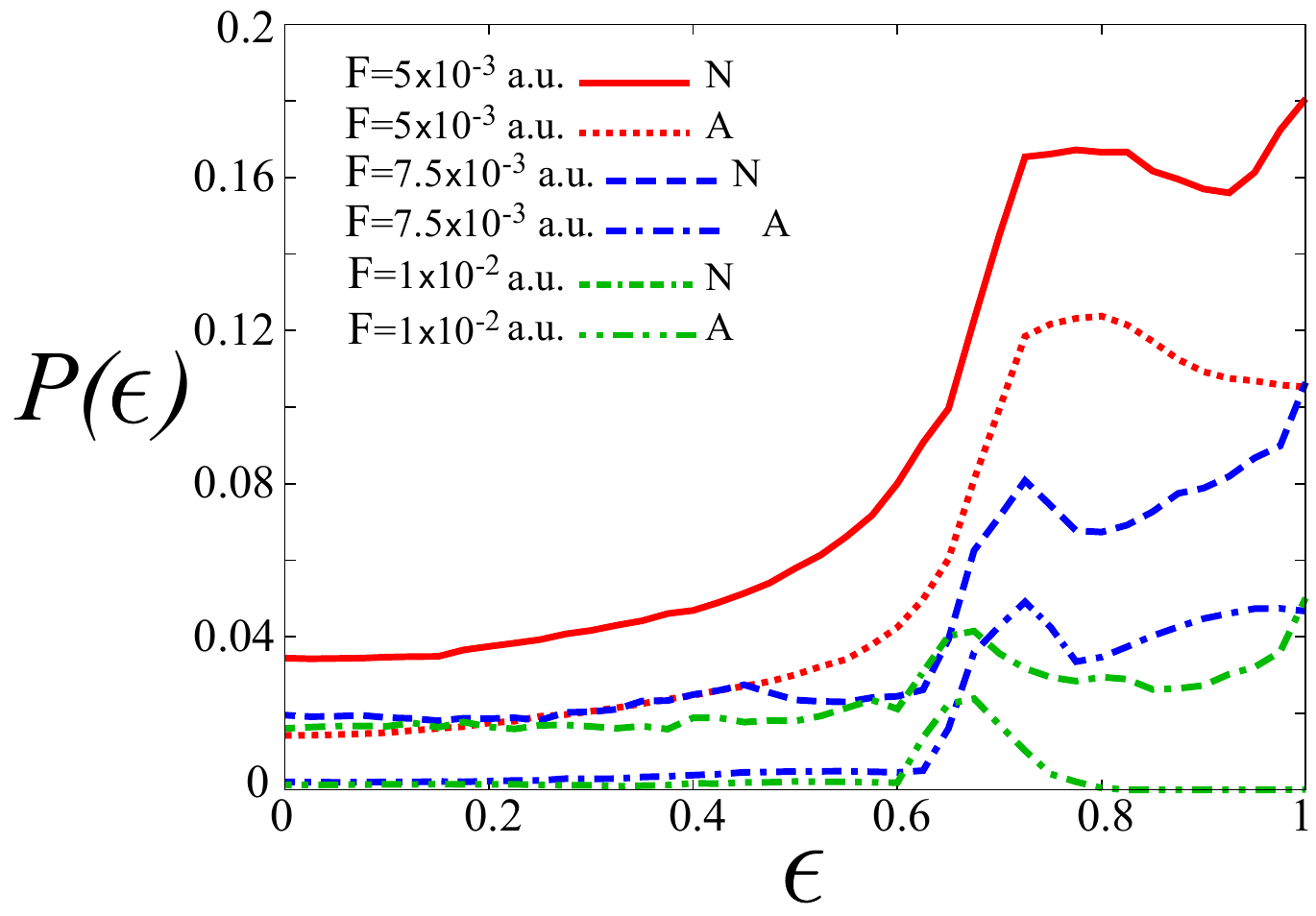}
 \caption{\label{fig:ProbaR200LargeF} Formation probabilities for  the (large)
laser intensities $F=5\times10^{-3}$ a.u., $F=7.5\times10^{-3}$ a.u.
and $F=1\times10^{-2}$ a.u.\ computed numerically (label N)
and with the approximate expression~\ref{eqn:EfLR} (label A).
In all cases the probabilities have been computed for an initial ensemble of
initial conditions with energy ${ E}_0=3\times 10^{-9}$.
The parameters of the pulse are $T_{\rm ru}=15~\mbox{ps}$, $T_{\rm p}=70~\mbox{ps}$ and $T_{\rm rd}=15~\mbox{ps}$. }
 \end{figure}

In Fig.~\ref{fig:paraF} we observe that when $F\gtrsim5\times10^{-3}$ a.u., there appears a
regime where $1< {\rm f} \ll {\rm f}^2$.
In this regime, the formation is very unlikely since the positive quadratic term in $\rm f$ is dominant for most of
the values of $(R_0,\phi_0)$, such that if ${\rm f}$ increases, the formation probability decreases.
However, there exist some particular values of
$(R_0,\phi_0)$ where this term can be made relatively small in comparison with the negative linear term in
$\rm f$. In this way, the minima of the quadratic term are obtained for $\phi_0=\pi/2 ~\mbox{mod } \pi$, and its value is proportional to
$$
{\rm min}_{\phi_0} g_2(\phi_0,\epsilon)= 4\left(\frac{1-2\epsilon^2}{1+\epsilon^2} \right)^2.
$$  
Therefore, for $\phi_0$ close to $\pi/2$ and for $R_0$ large enough, the quadratic term in
${\rm f}$ is the smallest one. We notice that it is even smaller for ellipticities close to $1/\sqrt{2}$. As a consequence, it is expected a higher formation probability for ellipticities close to $1/\sqrt{2}$, which explains the bump observed in the formation probability in Fig.~\ref{fi:probaR200}.
In Fig.~\ref{fig:ProbaR200LargeF}, we compare the evolution
of the formation probability $P(\epsilon)$
for the (large) laser intensities $F=5\times10^{-3}$ a.u., $F=7.5\times10^{-3}$ a.u.
and $F=1\times10^{-2}$ a.u.\ computed numerically
and by using the approximate expression~\ref{eqn:EfLR}. We notice that in all cases the approximate formation probabilities
display the same behavior as those numerically computed, with a pronounced increasing of formation around
$\epsilon\approx 1/\sqrt{2}$.

\subsubsection{For intermediate values of $\rm f$}
\begin{figure}
\includegraphics[width=0.6\textwidth]{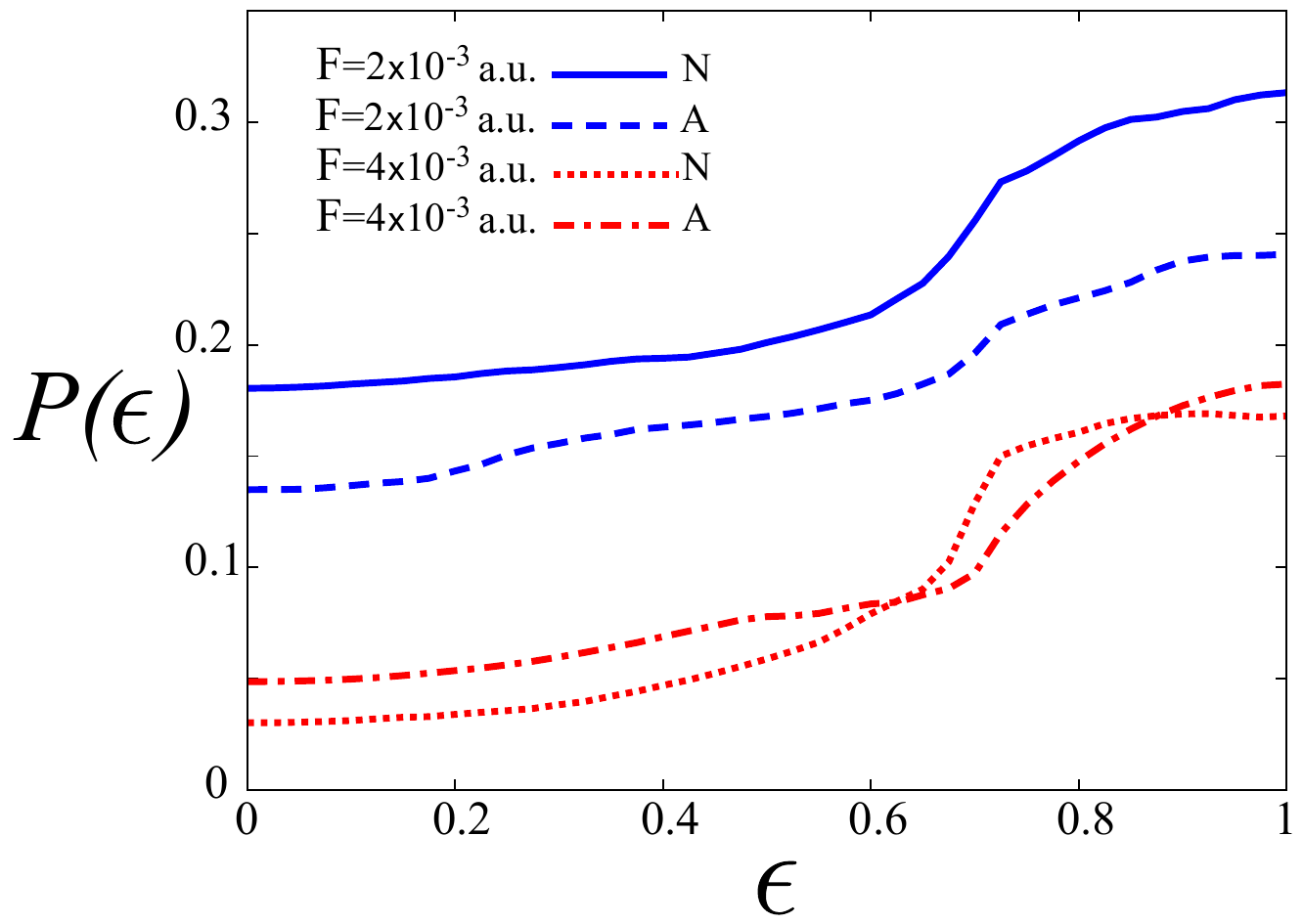}
 \caption{\label{fig:ProbaR200InterF} Formation probabilities for  the (intermediate)
laser intensities $F=2\times10^{-3}$ a.u. and $F=4\times10^{-3}$ a.u. calculated numerically (label N)
and with the approximate expression~\ref{eqn:EfLR} (label A).
In all cases the probabilities have been calculated for an initial ensemble of
initial conditions with energy ${ E}_0=3\times 10^{-9}$.
The parameters of the pulse are $T_{\rm ru}=15~\mbox{ps}$, $T_{\rm p}=70~\mbox{ps}$ and $T_{\rm rd}=15~\mbox{ps}$. }
 \end{figure}

When the electric field strength is in the range $1\times 10^{-3}$ a.u. $\lesssim F  \lesssim 5\times10^{-3}$ a.u.,  $\rm f$ and $\rm f^2$ are of the same order (see Fig.~\ref{fig:paraF}).
This regime is complex to analyze since all the three terms in Eq.~\ref{eqn:EfLR} compete.
For a fixed value of $\epsilon$ and taking into account that the effective field strength
parameter ${\rm f}$ is larger than unity for the considered field strength values, the
formation is expected to be enhanced for decreasing values
of the electric field strength $F$. Indeed, this is the observed behavior in the computations of $P(\epsilon)$
shown in Fig.~\ref{fig:ProbaR200InterF} for $F=2\times10^{-3}$ a.u. and
$F=4\times10^{-3}$ a.u.
Moreover, it is worth noticing in Fig.~\ref{fig:ProbaR200InterF} the good qualitative agreement between the evolution of the
formation probabilities $P(\epsilon)$ computed numerically and by using the approximate
expression Eq.~\ref{eqn:EfLR}. In order to analyze the increase of $P(\epsilon)$ for increasing values of
$\epsilon$ observed in
Fig.~\ref{fig:ProbaR200InterF}, we study the behavior of the approximate final energy $E_f$ given by
Eq.~\ref{eqn:EfLR}.
Two optimal situations appear for $\phi_0 \approx 0$ and $\phi_0 \approx \pi/2$ since in the neighborhood
of those directions, the induced momentum shift $\Delta P_\phi$ is approximately zero, such that
last term in Eq.~\ref{eqn:Ef} is negligible. Moreover, together with the conditions $P_{\phi}=0$, those directions are the invariant 
manifolds of the system along the minima $P_1$ ($\phi=0$) and the saddle points $P_2$ ($\phi=\pi/2$).
Along the directions of $P_{1,2}$, the induced momentum transfers $\Delta P_R$ are
\begin{eqnarray}
\label{eq:DPX_0}
(\Delta P_R)_{P_1} & = & -\frac{6  {\rm f}  \alpha_{\rm Rb}\alpha_{\rm Cs}}
{R^4} \left(\frac{2-\epsilon^2}{1+\epsilon^2}\right) , \\
\label{eq:DPX_Pi2}
(\Delta P_R)_{P_2} & = & -\frac{6 {\rm f}  \alpha_{\rm Rb}\alpha_{\rm Cs}}
{R^4} \left(\frac{2 \epsilon^2 -1}{1+\epsilon^2}\right).
\end{eqnarray}
Since $(\Delta P_R)_{P_1}$ is always negative, the final energy ${E}_f$ can be only be negative if
the initial conditions $P_R^0$ are positive. However,  $(\Delta P_R)_{P_2}$ is positive when $
\epsilon<1/\sqrt{2}$, negative when $\epsilon>1/\sqrt{2}$, and it takes zero value at $\epsilon=1/\sqrt{2}$.
Then, along the direction $P_2$ the final energy
${E}_f$ can be negative either for negative $P_R^0$ and $\epsilon<1/\sqrt{2}$ or for
positive $P_R^0$ and $\epsilon>1/\sqrt{2}$.
\begin{figure}
\centerline{\includegraphics[scale=.35]{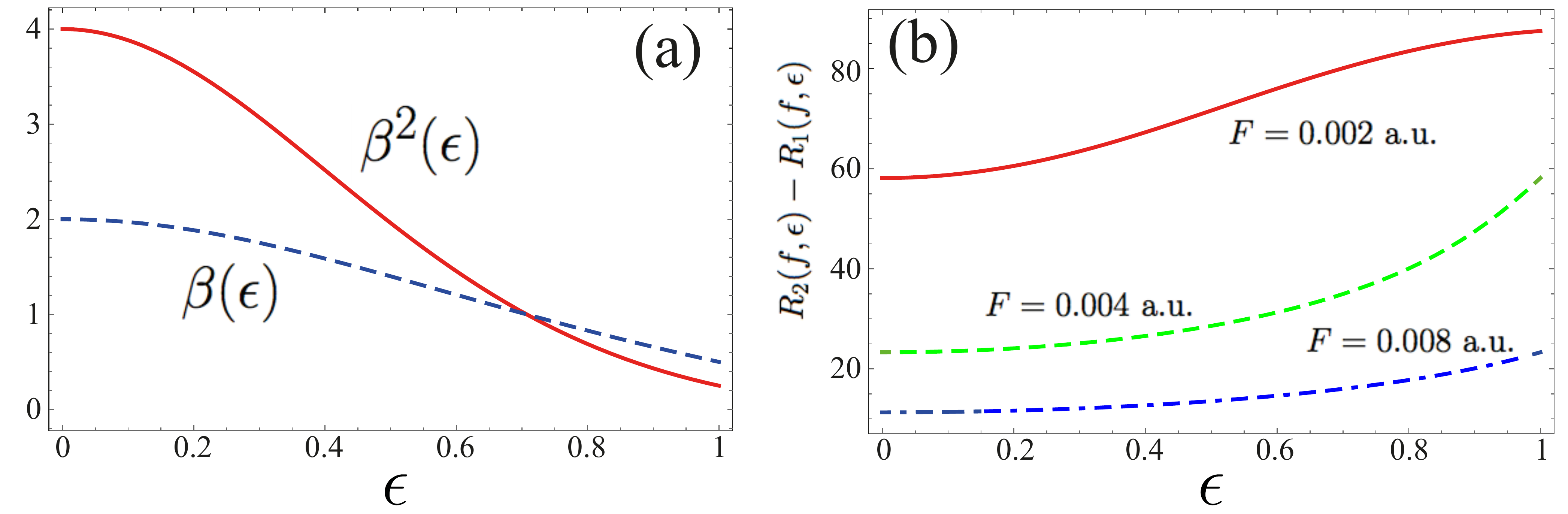}}
\caption{\label{fig:betas}(a) Evolution of the terms $\beta(\epsilon)$ and $\beta^2(\epsilon)$.
(b) Evolution of the distance $\Delta R_{2,1}(f,\epsilon)=
R_{2}(f,\epsilon)-R_{1}(f,\epsilon)$ between the two roots
of the final energy $({E}_f)_{P_1}$.}
\end{figure}

Along the direction $P_1$, the final energy~\ref{eqn:EfLR} becomes
\begin{equation}
\label{eqn:EfP1}
({E}_f)_{P_1}={E}_0-\frac{6  {\rm f}  \alpha_{\rm Rb}\alpha_{\rm Cs}}
{R^4} \ \sqrt{\frac{2}{\mu}\left({\cal E}_0+\frac{b_6}{R^6} \right)} \ \beta(\epsilon) 
+\frac{18  {\rm f}^2  \alpha_{\rm Rb}^2\alpha_{\rm Cs}^2}
{\mu R^8} \beta^2(\epsilon).
\end{equation}
\noindent
where $\beta(\epsilon)=g_1(\phi_0=0,\epsilon)=(2-\epsilon^2)/(1+\epsilon^2)>0$.
As we observe in Fig.~\ref{fig:betas}(a), the $\beta(\epsilon)$-terms involving the ellipticity are (positive) decreasing functions. Since the $\beta^2(\epsilon)$-term decreases faster than the $\beta(\epsilon)$-term such that for $\epsilon \ge 1/\sqrt{2}$, $\beta(\epsilon)\ge \beta^2(\epsilon)$, this qualitatively explains the increase of the formation probability for increasing ellipticity. Furthermore, when $R \longrightarrow \infty$, the function $({E}_f)_{P_1}$ tends to ${E}_0$, while
when $R \longrightarrow 0$, $({E}_f)_{P_1}$ tends to $+\infty$. The function $({E}_f)_{P_1}$
has two roots $R_1(f,\epsilon)$ and $R_2(f,\epsilon)$ such that $R_1(f,\epsilon)\le R_2(f,\epsilon)$.
In the interatomic region between these two roots, $({E}_f)_{P_1}$ is negative, and as a consequence, there is formation when $R_0$ is between
$R_1(f,\epsilon)$ and $R_2(f,\epsilon)$. In Fig.~\ref{fig:betas}(b),
$\Delta R_{2,1}(f,\epsilon)=R_{2}(f,\epsilon)-R_{1}(f,\epsilon)$ is represented as a function of $\epsilon$ for three fixed values of $F$.
In all cases, the distance $\Delta R_{2,1}(f,\epsilon)$ increases with $\epsilon$ and
decreases with $F$.

Along the direction $P_2$, the final energy \ref{eqn:EfLR} takes the form
\begin{equation}
\label{eqn:EfP2}
({E}_f)_{P_2}={E}_0 - \frac{6 \ {\rm f} \ \alpha_{\rm Rb}\alpha_{\rm Cs}}
{R^4} \ \sqrt{\frac{2}{\mu}\left({\cal E}_0+\frac{b_6}{R^6} \right)} \ \gamma(\epsilon) 
+\frac{18 \ {\rm f}^2 \ \alpha_{\rm Rb}^2\alpha_{\rm Cs}^2}
{\mu R^8} \gamma(\epsilon)^2,
\end{equation}
\noindent
where $\gamma(\epsilon)=g_1(\phi_0=\pi/2,\epsilon)/2=(2\epsilon^2-1)/(1+\epsilon^2)$.
In particular, when $\epsilon=1/\sqrt{2}$, the
factor $\gamma(\epsilon)$ vanishes
and formation is not possible along the direction $P_2$.
\begin{figure}
\centerline{\includegraphics[scale=.35]{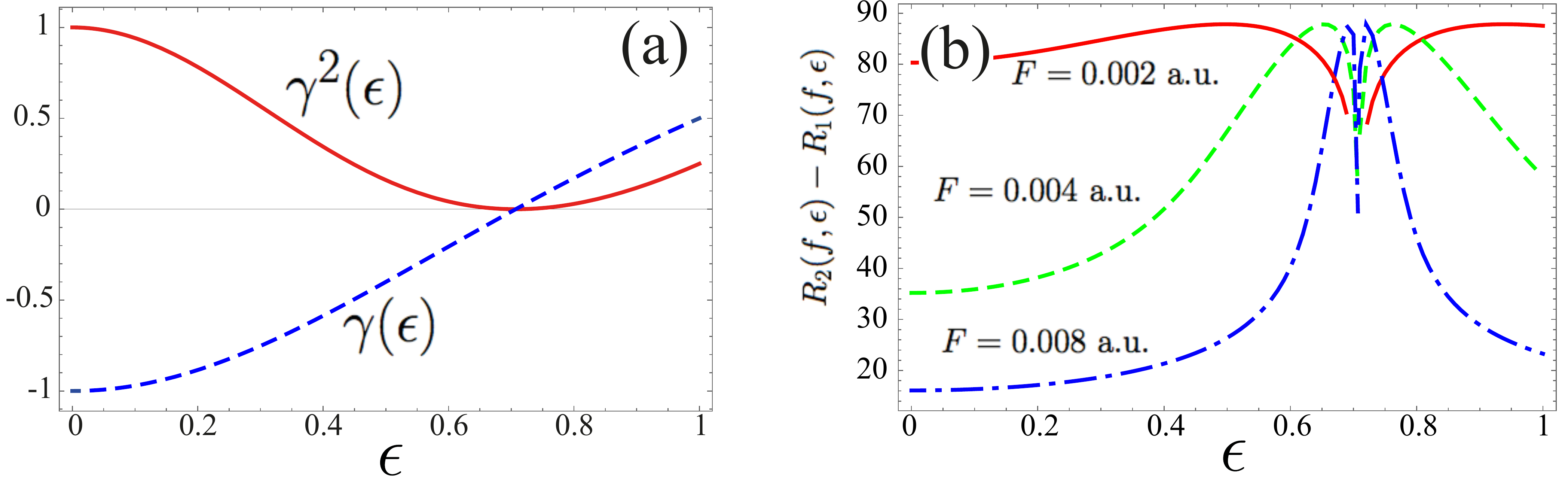}}
\caption{(a) Evolution of the terms $\gamma(\epsilon)$ and $\gamma^2(\epsilon)$.
(b) Evolution of the distance $\Delta R_{2,1}(f,\epsilon)=
R_{2}(f,\epsilon)-R_{1}(f,\epsilon)$ between the two roots
of the final energy $({E}_f)_{P_2}$.}
\label{fig:gammas}
\end{figure}
The evolution of the factors $\gamma(\epsilon)$ and $\gamma^2(\epsilon)$
is depicted in Fig.~\ref{fig:gammas}(a).
In both intervals $0\le \epsilon <1/\sqrt{2}$ and $1/\sqrt{2} < \epsilon \le 1$, the
function $({E}_f)_{P_2}$
has two roots $R_1(f,\epsilon)$ and $R_2(f,\epsilon)$ such that $R_1(f,\epsilon)\le R_2(f,\epsilon)$.
Therefore, we follow again the evolution of the distance
$\Delta R_{2,1}(f,\epsilon)=R_{2}(f,\epsilon)-R_{1}(f,\epsilon)$ between these roots as a function of the ellipticity in each of these intervals, and
the result of this study is shown in Fig.~\ref{fig:gammas}(b).
We observe that, in all cases, when the ellipticity approaches
the critical value $1/\sqrt{2}$, the distance $\Delta R_{2,1}(f,\epsilon)$ sharply tends to 
zero because at
$\epsilon =1/\sqrt{2}$, we have $({E}_f)_{P_2}={E}_0>0$.
On the other hand, Fig.~\ref{fig:gammas}(b) shows that the distance between the roots
increases for increasing ellipticity in the
interval $0\le \epsilon <1/\sqrt{2}$, while it decreases in the interval $1/\sqrt{2} < \epsilon \le 1$.
\begin{figure}
\includegraphics[width=0.6\textwidth]{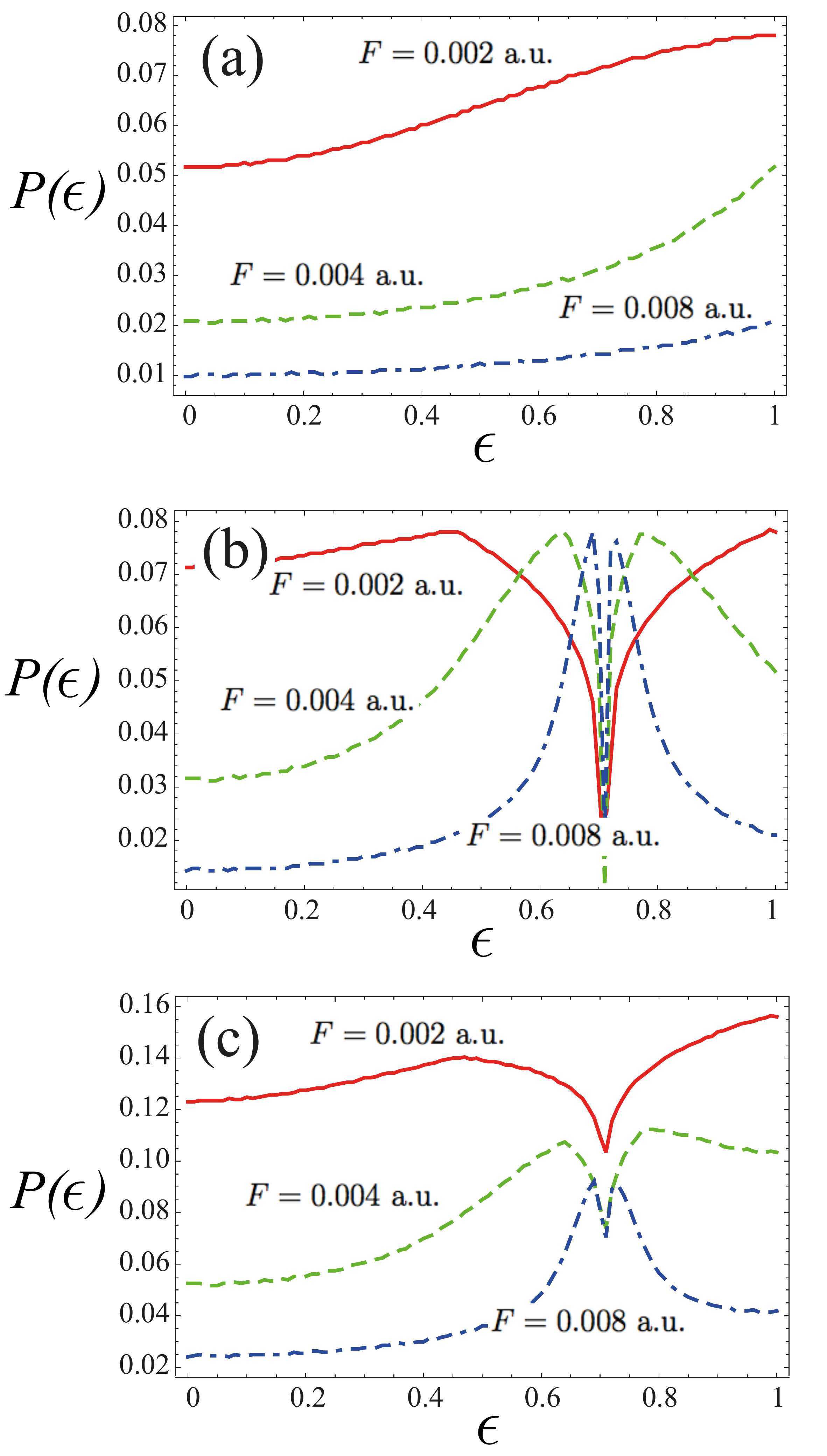}
 \caption{Panels (a) and (b): Formation probability obtained, respectively,
 with Eq.~\ref{eqn:EfP1} and with Eq.~\ref{eqn:EfP2}
 as a function of ellipticity $\epsilon$ for $F=2 \times 10^{-3}$ a.u. (solid red line), $F=4 \times 10^{-3}$ a.u. (dashed green line)
and $F=8 \times 10^{-3}$ a.u. (dashed-dotted blue line). c) Joint probability calculated with
Eqs.~\ref{eqn:EfP1}-\ref{eqn:EfP2}. The probabilities have been calculated for an initial ensemble of
initial conditions with energy ${ E}_0=3\times 10^{-9}$ a.u.
The parameters of the pulse are $T_{\rm ru}=15~\mbox{ps}$, $T_{\rm p}=70~\mbox{ps}$ and $T_{\rm rd}=15~\mbox{ps}$.}
\label{fig:ProbabilitiesPHI} 
 \end{figure}
We use the final energies $({E}_f)_{P_1}$ and $({E}_f)_{P_2}$ to compute
numerically the respective formation probabilities. To do that, we use an ensemble of initial
conditions with $P_{\phi}=0$,  $\phi=0$ for $({E}_f)_{P_1}$,
$\phi=\pi/2$ for $({E}_f)_{P_1}$ and random $R(0) \in [6.2329, 200]$ a.u. The results
of these computations are shown in Fig.~\ref{fig:ProbabilitiesPHI}.
In Fig.~\ref{fig:ProbabilitiesPHI}(a) the evolution of the formation probability is shown
using the final energy given by Eq.~\ref{eqn:EfP1}. As expected, the probability mimics
the behavior of $\Delta R_{2,1}(f,\epsilon)$ shown in Fig.~\ref{fig:betas}(b). Indeed, the
formation probability increases for
increasing ellipticity and it decreases for increasing electric fields.
The behavior of the formation probability along the direction of the saddle point $P_2$
given by the final energy Eq.~\ref{eqn:EfP2} also follows the pattern of  $\Delta R_{2,1}(f,\epsilon)$
shown in Fig.~\ref{fig:gammas}(b). Note that, at the critical ellipticity $\epsilon=1/\sqrt{2}$, the
formation probability is zero. Finally, when the joint formation probability along the
directions of $P_{1,2}$ is calculated
[see Fig.~\ref{fig:ProbabilitiesPHI}(c)], except in the neighborhood of $\epsilon=1/\sqrt{2}$, 
its behavior resembles the one obtained in Fig.~\ref{fig:ProbaS4Dell}.
Roughly speaking, the global behavior of the formation probability is made of two
main contributions: one of them coming from the direction $P_1$ and the other one
from the direction $P_2$.
\begin{figure}
\includegraphics[width=0.6\textwidth]{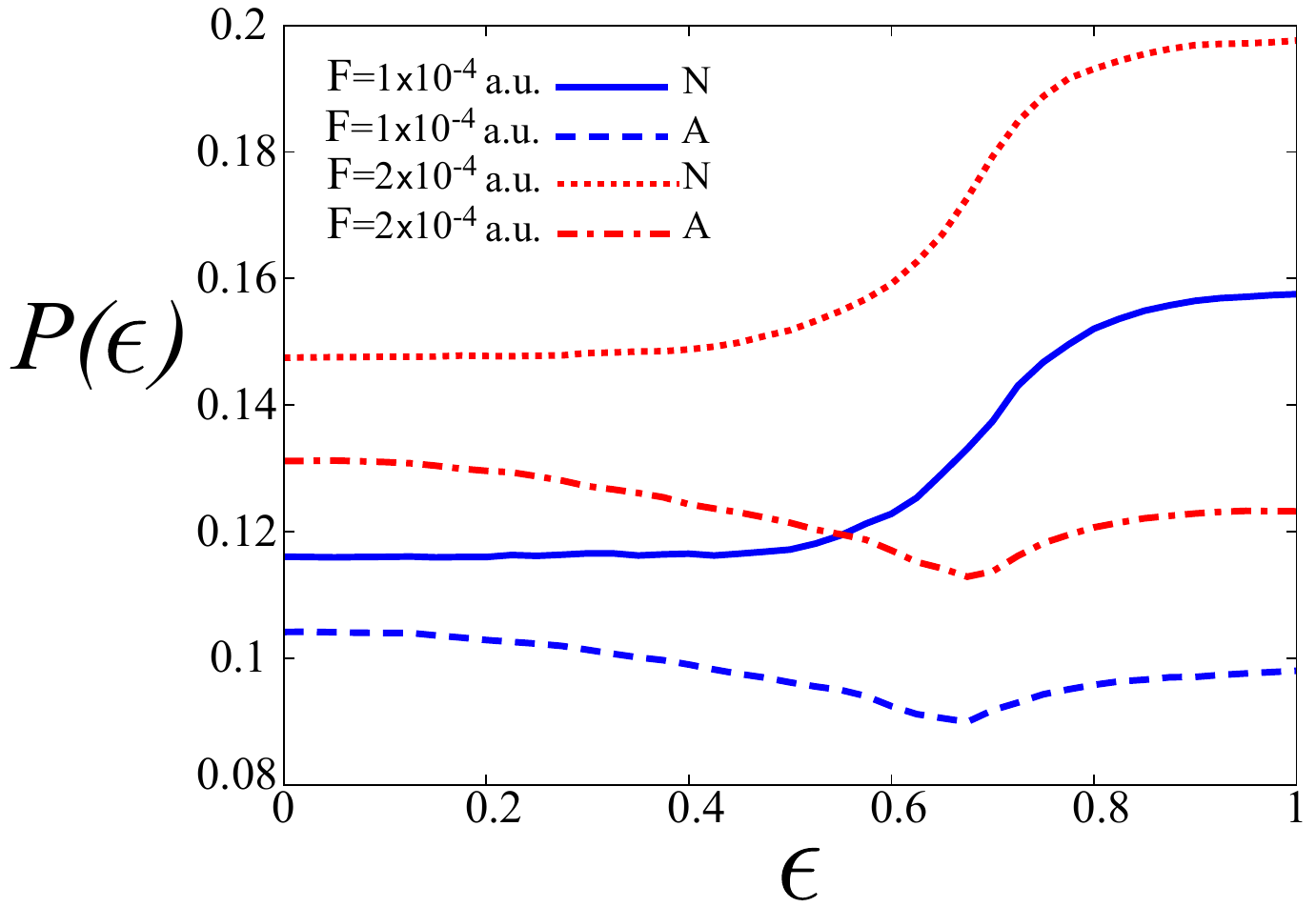}
 \caption{\label{fig:ProbaR200SmallF} Formation probabilities for  the (small)
laser intensities $F=1\times10^{-4}$ a.u. and $F=2\times10^{-4}$ a.u.\ calculated numerically (label N)
and with the approximate expression~\ref{eqn:EfLR} (label A).
In all cases, the probabilities have been calculated for an initial ensemble of
initial conditions with energy ${ E}_0=3\times 10^{-9}$ a.u.
The parameters of the pulse are $T_{\rm ru}=15~\mbox{ps}$, $T_{\rm p}=70~\mbox{ps}$ and $T_{\rm rd}=15~\mbox{ps}$. }
 \end{figure}

\subsubsection{For small values of $\rm f$}
For relatively small values of the field strength below $F \lesssim 1\times10^{-3}$ a.u. (see Fig.\ref{fig:paraF}), it is possible
to assume that $\rm f >> {\rm f}^2$, such that the condition \ref{eqn:EfLR} for formation becomes
$$
3 \ \sqrt{\frac{2}{\mu}\left({E}_0+\frac{b_6}{R_0^6}\right)}\frac{\alpha_{\rm Rb}\alpha_{\rm Cs}}{R_0^4}
 \left\vert g_1(\phi_0,\epsilon)\right\vert > \frac{E_0}{{\rm f}}.
$$
We readily see that if ${\rm f}$ increases, the threshold for formation $E_0/{\rm f}$ decreases, and hence the formation probability $P(\epsilon)$ increases for all values of the ellipticity. Furthermore, we notice that the function $\vert g_1(\cdot,\epsilon)\vert$,
displays two distinct behaviors, one for $\epsilon<1/\sqrt{2}$ and one for $\epsilon>1/\sqrt{2}$.
For $\epsilon<1/\sqrt{2}$, the function $ \vert g_1(\cdot,\epsilon)\vert$ has a maximum at $\phi_0=0$ with a value of $2(2-\epsilon^2)/(1+\epsilon^2)$, a local maximum at $\phi_0=\pi/2$ with a value of $2(1-2\epsilon^2)/(1+\epsilon^2)$, and vanishes for an intermediate value of $\phi_0$. As $\epsilon$ is increased up to $1/\sqrt{2}$, the maximum and the local maximum are decreasing. For a given value of $R_0$, only a finite set of $\phi_0$ leads to formation. Therefore, a decreasing of the formation probability with $\epsilon$ is expected. 
At $\epsilon=1/\sqrt{2}$, the local maximum and the minimum merge, such that for
$\epsilon\geq 1/\sqrt{2}$, the function $\vert g_1(\cdot,\epsilon)\vert$ presents a maximum at $\phi_0=0$ with a value of $2(2-\epsilon^2)/(1+\epsilon^2)$ and a minimum at $\phi_0=\pi/2$ with a value of $2(2\epsilon^2-1)/(1+\epsilon^2)$. As $\epsilon$ increases from $1/\sqrt{2}$, the maximum decreases, but the minimum is increasing. In this range of ellipticities, for a value of $R_0$ sufficiently small (below a value related to the minimum of $\vert g_1(\cdot,\epsilon)\vert$), all values of $\phi_0$ lead to formation. Therefore, the formation probability would tend to increase with ellipticity.

However, as we already observed in Fig.~\ref{fig:ProbaS4Dell}, the
above described behavior of $P(\epsilon)$ is not the scenario
observed when $P(\epsilon)$ is numerically computed for small values of $\rm f$.
For example, in Fig.~\ref{fig:ProbaR200SmallF}, we compare the evolution of $P(\epsilon)$ computed numerically with the approximate one given by Eq.~\ref{eqn:EfLR} for the (small) field values $F=1\times10^{-4}$ a.u. and $F=2\times10^{-4}$ a.u. Indeed, while in both cases the approximate formation probability follows the above described behavior, the numerical
formation probabilities show a monotonic increase for increasing values of the ellipticity $\epsilon$.
In this way, for small enough values of $F$, there is a clear disagreement
between the numerical and the approximate results.
The main reason of this disagreement  for small values of $F$ is that, during the
pulse, the molecule-laser interaction is not strong enough to be the dominant
interaction between the two atoms. Then, the approximation $d{\cal E}(R)/dR\approx 0$ leading to the radial momentum transfer $\Delta P_R$ given by Eq.~\ref{eq:transfer1} is only valid for large values of $R_0$, such that the static approximation fails .

\section{Conclusions}
\label{Conclusions}
Under the Born-Oppenheimer approximation, we analyzed the classical
rovibrational dynamics and the formation probability of the
alkali polar molecule RbCs in its electronic ground state in the presence of a relatively long
elliptically polarized laser pulse. After the average of the dynamics over the
(fast) frequency of the laser, we obtain the Hamiltonian of the system where only the interaction between the laser field
and the molecular polarizability is left.
Although the resulting Hamiltonian has 3$+1/2$ degrees of freedom, the system possesses an
invariant manifold where the dynamics is confined to the polarization plane of the laser field.
In this way, our investigation is restricted to that invariant manifold where the system has
2$+1/2$ degrees of freedom. Furthermore, besides the energy of the system, the dynamics depends on the
laser pulse parameters such as the electric field strength $F$, the pulse duration and its ellipticity $\epsilon$.

We construct an analytical representation of the potential energy surface by fitting the corresponding
available data  for the RbCs of the potential energy curve and of the parallel and perpendicular
components of the molecular polarizability. In this way, the landscape of the potential energy
surface of the problem during the plateau of the pulse is analyzed by
studying the evolution of the number and the stability of its critical points as a function of the electric field strength and the ellipticity of the laser.  When the laser pulse is linearly polarized, $\epsilon=0$, in the short 
range of interatomic distances, there are two minima and two saddle points, while in the long range,
there are two maxima. For increasing values of $\epsilon$, the maxima move away off the saddles and
the minima, such that at $\epsilon \approx 1/\sqrt{2}$ they disappear. At the same time, the depth
of the potential wells determined by the minima and the saddle points decreases in such a way that, at
$\epsilon=1$ they come into coincidence and they disappear.

By means of Poincar\'e surfaces of section, we have also studied the evolution of the
phase space structures of the problem during the plateau and for
an energy value which always remains around the
dissociation threshold. By using different field strengths $F$ and for increasing ellipticity, we
find that for small values of $\epsilon$, the phase space of the system resembles a chaotic
pendulum. As expected, when the ellipticity tends to 1, the system
approaches its integrable limit, and the
chaotic regions as well as the pendulum-like phase space structure
disappear. We find that, when the considered (fixed) energy is above the dissociation threshold, most of
the chaotic orbits rapidly dissociate.

Our numerical computations showed that the formation process of the dimer is very sensitive to the
laser field parameters such as the electric field strength, the pulse duration and the ellipticity.
In order to elucidate the influence of these parameters on the formation process, we use
a static assumption. In this way, we assume that during the duration of the pulse both the
interatomic distance $R$ and the $\phi$ do not change significantly, such that the formation
is caused by small changes in the momenta $P_R$ and $P_{\phi}$. 
Using the asymptotic values of the Silberstein expressions of the polarizabilities, we obtain
an approximate expression for the final energy $E_f$ of the dimer after the pulse.
This approximate expression of $E_f$ allows us to elucidate the strong influence of the ellipticity in the formation probability, notably with the presence of a critical ellipticity around $1/\sqrt{2}$.

\end{document}